\documentclass[floatfix,a4paper,pra,amsmath,amssymb,aps,twocolumn,10pt]{revtex4-1}

\usepackage[matrix,frame,arrow]{xy}
\usepackage{amsmath}
\newcommand{\qw}[1][-1]{\ar @{-} [0,#1]}
\newcommand{\qwx}[1][-1]{\ar @{-} [#1,0]}
\newcommand{\cw}[1][-1]{\ar @{=} [0,#1]}

\newcommand{\gate}[1]{*{\xy *+<.6em>{#1};p\save+LU;+RU **\dir{-}\restore\save+RU;+RD **\dir{-}\restore\save+RD;+LD **\dir{-}\restore\POS+LD;+LU **\dir{-}\endxy} \qw}
\newcommand{\meter}{\gate{\xy *!<0em,1.1em>h\cir<1.1em>{ur_dr},!U-<0em,.4em>;p+<.5em,.9em> **h\dir{-} \POS <-.6em,.4em> *{},<.6em,-.4em> *{} \endxy}}

\newcommand{\control}{*!<0em,.025em>-=-{\bullet}}

\newcommand{\ctrl}[1]{\control \qwx[#1] \qw}

\newcommand{\targ}{*!<0em,.019em>=<.79em,.68em>{\xy {<0em,0em>*{} \ar @{ - } +<.4em,0em> \ar @{ - } -<.4em,0em> \ar @{ - } +<0em,.36em> \ar @{ - } -<0em,.36em>},<0em,-.019em>*+<.8em>\frm{o}\endxy} \qw}
\newcommand{\qswap}{*=<0em>{\times} \qw}
\newcommand{\Qcircuit}[1][0em]{\xymatrix @*[o] @*=<#1>}

\usepackage[british]{babel}
\usepackage{hyperref}
\usepackage{graphicx}
\usepackage{subfigure}

\usepackage{listings}
\lstdefinelanguage{qpc}
  {keywords={for, to, if, then, while, do},
    sensitive=true,
    comment=[l]{C:},
    morestring=[b]"
  }

\lstdefinelanguage{qcl}
  {keywords={operator, procedure, qureg, xor, reset, measure, print, until, int,
  real, for, to, step, if, else, dump, const, quconst, quvoid, cond, qufunct,
  extern,quscratch},
    sensitive=true,
    morecomment=[l]{//},
    morecomment=[s]{/*}{*/},
    morestring=[b]"
  }

\lstdefinelanguage{cqpl}
  {keywords={module, new, qbit, qint, if, then, receive, from, print, dump,
  bit, call, send, to, measure, proc, in, then, skip, else, while, int,proc},
    sensitive=true,
    morecomment=[l]{//},
    morecomment=[s]{/*}{*/},
    morestring=[b]",
  }

\lstdefinelanguage{lanq}
    {morekeywords={channelEnd, channel, withends, send, recv, int, bit, qint,
    void, dump_q, measure, else, if, then, aliasfor, fork, new, while, return,
    qbit, qtrit, boolean},
    sensitive=true,
    morecomment=[l]{//},
    morecomment=[s]{/*}{*/},
    morestring=[b]",
  }
  
\newcommand{\eg}{\emph{e.g.}}
\newcommand{\ie}{\emph{i.e.}}

\newcommand{\ket}[1]{\ensuremath{|#1\rangle}}
\newcommand{\bra}[1]{\ensuremath{\langle#1|}}
\newcommand{\ketbra}[2]{\ensuremath{\ket{#1}\bra{#2}}}
\newcommand{\proj}[1]{\ensuremath{\ketbra{#1}{#1}}}
\newcommand{\Proj}[1]{\proj{#1}}

\newcommand{\Operators}[1]{\Operators{#1}}
\newcommand{\HS}[1]{\ensuremath{\mathcal{#1}}} 

\newcommand{\Id}{\ensuremath{\mathbb{I}}}

\newcommand{\Cplx}{\ensuremath{\mathbb{C}}}

\newcommand{\Z}{\ensuremath{\mathbb{Z}}}
\newcommand{\set}[2]{\ensuremath{\left\{#1|#2\right\}}}
\newcommand{\re}[1]{\ensuremath{\mathrm{Re}\left(#1\right)}}
\newcommand{\im}[1]{\ensuremath{\mathrm{Im}\left(#1\right)}}

\newcommand{\BigOh}[1]{\ensuremath{O(#1)}}
\newcommand{\complexity}[1]{\ensuremath{\mathbf{#1}}}

\newtheorem{theorem}{Theorem}

\newtheorem{definition}{Definition}

\newtheorem{problem}{Problem}

\newtheorem{hypo}{Hypothesis}

\newcommand{\yes}{\checkmark}
\newcommand{\no}{--}

\begin{document}

\title{Models of quantum computation and quantum programming languages}

\author{Jaros{\l}aw Adam MISZCZAK}

\affiliation{Institute of Theoretical and Applied
Informatics,\\ Polish Academy of Sciences, Ba{\l}tycka 5, 44-100 Gliwice,
Poland}

\begin{abstract}
The goal of the presented paper is to provide an introduction to the basic
computational models used in quantum information theory. We review various
models of quantum Turing machine, quantum circuits and quantum random access
machine (QRAM) along with their classical counterparts. We also provide an
introduction to quantum programming languages, which are developed using the
QRAM model. We review the syntax of several existing quantum programming
languages and discuss their features and limitations.
\end{abstract}

\maketitle

\section{Introduction}
Computational process must be studied using the fixed model of computational
device. This paper introduces the basic models of computation used in quantum
information theory. We show how these models are defined by extending classical
models.

We start by introducing some basic facts about classical and quantum Turing
machines. These models help to understand how useful quantum computing can be.
It can be also used to discuss the difference between quantum and classical
computation. For the sake of completeness we also give a brief introduction to
the main results of quantum complexity theory. Next we introduce Boolean
circuits and describe the most widely used model of quantum computation, namely
quantum circuits. We focus on this model since many presented facts about
quantum circuits are used in the following sections. Finally we introduce
another model which is more suited for defining programming languages operating
on quantum memory --- quantum random access machine (QRAM).

We also describe selected examples of the existing quantum programming
languages. We start by formulating the requirements which must be fulfilled by
any universal quantum programming language. Next we describe languages based on
imperative paradigm -- QCL (Quantum Computation Language) and LanQ. We also
describe recent research efforts focused on implementing languages based on
functional paradigm and discuss the advantages of a language based on this
paradigm. As the example of functional quantum programming language we present
cQPL.

We introduce the syntax and discuss the features of the presented languages. We
also point out their weaknesses. For the sake of completeness a few examples of
quantum algorithms and protocols are presented. We use these examples to
introduce the main features of the presented languages.

Note that we will not discuss problems related to the physical realisation of
the described models. We also do not cover the area of quantum error correcting
codes, which aims to provide methods for dealing with decoherence in quantum
systems. For an introduction to these problems and recent progress in this area
see \eg~\cite{bouwmeester00physics,ladd10quantumcomputers}.

One should be also aware that the presented overview of existing models of
quantum computation is biased towards the models interesting for the development
of quantum programming languages. Thus we neglect some models which are not
directly related to this area (\eg\ quantum automata or topological quantum
computation). 

\subsection{Quantum information theory}
Quantum information theory is a new, fascinating field of research which aims to
use quantum mechanical description of the system to perform computational tasks.
It is based on quantum physics and classical computer science, and its goal is
to use the laws of quantum mechanics to develop more powerful algorithms and
protocols.

\begin{figure}[ht]
    \centering
    \includegraphics[width=\columnwidth]{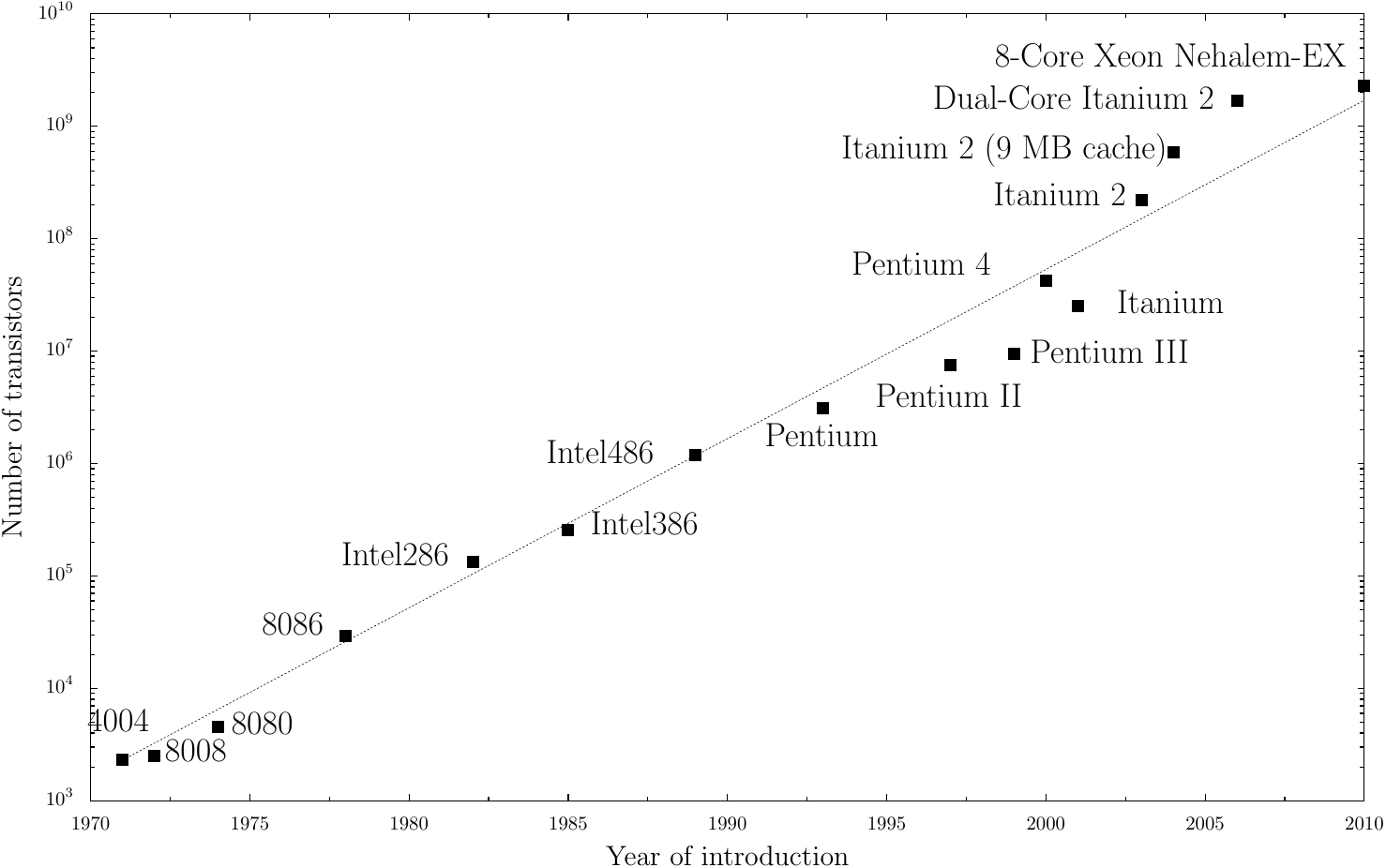}
    \caption{Illustration of Moore's hypothesis. The number of transistors which
    can be put on a single chip grows exponentially. The squares represent
    microprocessors introduced by Intel
    Corporation~\protect\cite{intelmooreslaw}. The dotted line illustrates the
    rate of growth, with the number of transistors doubling every two years.}
    \label{fig:moores-law}
\end{figure}

According to the Moore's Law \cite{moore65cramming,intelmooreslaw} the number of
transistors on a given chip is doubled every two years (see
Figure~\ref{fig:moores-law}). Since classical computation has its natural
limitations in the terms of the size of computing devices, it is natural to
investigate the behaviour of objects in micro scale.

Quantum effects cannot be neglected in microscale and thus they must be taken
into account when designing future computers. Quantum computation aims not only
at taking them into account, but also at developing methods for controlling
them. Quantum algorithms and protocols are recipes how one should control
quantum system to achieve higher efficiency.

Information processing on quantum computer was first mentioned in 1982 by
Feynman~\cite{feynman82simulating}. This seminal work was motivated by the fact
that simulation of a quantum system on the classical machine requires
exponential resources. Thus, if we could control a physical system at the
quantum level we should be able to simulate other quantum systems using such
machines.

The first quantum protocol was proposed two years later by Bennett and
Brassard~\cite{bb84}. It gave the first example of the new effects which can be
obtained by using the rules of quantum theory for information processing. In
1991 Ekert described the protocol~\cite{ekert91bell} showing the usage of
quantum entanglement~\cite{epr} in communication theory.

Today we know that thanks to the quantum nature of photons it is possible to
create unconditionally secure communication links~\cite{boudaPHD} or send
information with the efficiency unachievable using classical carriers. During
the last few years quantum cryptographic protocols have been implemented in
real-world systems. Quantum key distribution is the most promising application
of quantum information theory, if one takes practical
applications~\cite{ursin07144,secoqo} into account.

On the other hand we know that the quantum mechanical laws of nature allow us to
improve the solution of some
problems~\cite{shor94algorithms,shor97algorithms,grover97haystack}, construct
games~\cite{eisert99games,meyer00games} and random
walks~\cite{kempe03overview,kosik03models} with new properties.

Nevertheless, the most spectacular achievements in quantum information theory up
to the present moment are: the quantum algorithm for factoring numbers and
calculating discrete logarithms over finite field proposed in the late nineties
by Shor~\cite{shor97algorithms,shor97algorithms}. The quantum algorithm solves
the factorisation problem in polynomial time, while the best known probabilistic
classical algorithm runs in time exponential with respect to the size of input
number. Shor's factorisation algorithm is one of the strongest arguments for the
conjecture that quantum computers can be used to solve in polynomial time
problems which cannot be solved classically in reasonable (\ie\ polynomial)
time.

Taking into account research efforts focused on discovering new quantum
algorithms it is surprising that for the last ten years no similar results have
been obtained~\cite{shor04progress,lomonaco05search}. One should note that
there is no proof that quantum computers can actually solve
\complexity{NP}-complete problems in polynomial
time~\cite{bernstein97complexity,fortnow03one}. This proof could be given by
quantum algorithms solving in polynomial time problems known to be
\complexity{NP}-complete such as $k$-colorability. The complexity of quantum
computation remains poorly understood. We do not have much evidence how useful
quantum computers can be. Still much remains to be discovered in the area of the
relations between quantum complexity classes such as \complexity{BQP} and
classical complexity classes like~\complexity{NP}.

\subsection{Progress in quantum algorithms}
Due to the slow progress in discovering new quantum algorithms novel methods for
studying the impact of quantum mechanics on algorithmic problems were proposed.

The first of these methods aims at applying the rules of quantum mechanics to
game theory \cite{meyer00games,klarreich01playing}. Classical games are used to
model the situation of conflict between competing agents. The simplest
application of quantum games is presented in the form of quantum prisoners
dilemma \cite{eisert99games}. In this case one can analyse the impact of quantum
information processing on classical scenarios. On the other hand quantum games
can be also used to analyse typical quantum situations like state estimation and
cloning~\cite{lee03cloning}. 

Quantum walks provide the second promising method for developing new quantum
algorithms. Quantum walks are the counterparts of classical random walks
\cite{kempe03overview,kosik03models}. For example, in
\cite{ambainis07distinctness} the quantum algorithm for element distinctness
using this method was proposed. It requires $\BigOh{n^{2/3}}$ queries to
determine if the input $\{x_1,\ldots,x_n\}$ consisting of $n$ elements contains
two equal numbers. Classical algorithm solving this problem requires
$\BigOh{n\log n}$ queries. The generalisation of this algorithm, with
applications to the problem of subset finding, was described in
\cite{childs05subset}. Other application of quantum walks include searching
algorithms \cite{childs04spatial} and subset finding problem. It was also shown
that quantum walks can be used to perform a universal quantum computation
\cite{childs09universal,hines07gates}. In \cite{ambainis03algorithmic} the
survey of quantum algorithms based on quantum walks is presented. More
information concerning recent developments in quantum walks and their
applications can be found in~\cite{santha08walkbased}. 

One should note that the development of quantum algorithms is still a very
lively area of research~\cite{lomonaco05search,mscs-special-intro}. General
introduction to quantum algorithms can be found in~\cite{mosca11algorithms}. The
in-depth review of the recent results in the area of quantum algorithms for
algebraic problems can be found in~\cite{childs10algorithms}. 

\section{Computability}
Classically computation can be described using various models. The choice of
the model used depends on the particular purpose or problem. Among the most
important models of computation we can point:
\begin{itemize}
    \item {\bf Turing Machine} introduced in 1936 by Turing and used as the main
    model in complexity theory~\cite{papadimitriou}.

    \item {\bf Random Access Machine}
    \cite{cook73time,shepherdson63computability} which is the example of
    register machines; this model captures the main features of modern computers
    and provides a theoretical model for programming languages.

    \item {\bf Boolean circuits} \cite{vollmer} defined in terms of logical
    gates and used to compute Boolean functions $f:\{0,1\}^m\mapsto\{0,1\}^n$;
    they are used in complexity theory to study circuit complexity.

    \item {\bf Lambda calculus} defined by Church \cite{church36unsolvable} and
    used as the basis for many functional programming languages \cite{sicp}.

    \item {\bf Universal programming languages} which are probably the most
    widely used model of computation \cite{mitchell03concepts}.
\end{itemize}

It can be shown that all these models are equivalent
\cite{papadimitriou,vollmer}. In other words the function which is computable
using one of these models can be computed using any other model. It is quite
surprising since Turing machine is a very simple model, especially when compared
with RAM or programming languages.

In particular the model of a multitape Turing machine is regarded as a canonical
one. This fact is captured by the Church-Turing hypothesis.

\begin{hypo}[Church-Turing]
Every function which would be naturally regarded as computable can be
computed by a universal Turing machine.
\end{hypo}

Although stated as a hypothesis, this thesis is one of the fundamental axioms of
modern computer science. A Universal Turing machine is a machine which is able
to simulate any other machine. The simplest method for constructing such device
is to use the model of a Turing machine with two tapes~\cite{papadimitriou}.

Research in quantum information processing is motivated by the extended
version of Church-Turing thesis formulated by Deutsch \cite{deutsch85principle}.

\begin{hypo}[Church-Turing-Deutsch]\label{hyp:ctd}
Every physical process can be simulated by a universal computing device.
\end{hypo}

In other words this thesis states that if the laws of physics are used to
construct a Turing machine, this model might provide greater computational
power when compared with the classical model. Since the basic laws of physics 
are formulated as quantum mechanics, this improved version of a Turing machine
should be governed by the laws of quantum physics.

In this section we review some of these computational models focusing on their
quantum counterparts. The discussion of quantum programming languages, which
are based on the quantum random access machines (QRAM), is presented
in Section \ref{sec:langs}.

We start be recalling the basic facts concerning a Turing machine. This model
allows to establish clear notion of computational resources like time and space
used during computation. It is also used to define other models introduced in
this section precisely.

On the other hand for practical purposes the notion of Turing machine is clumsy.
Even for simple algorithms it requires quite complex description of transition
rules. Also, programming languages defined using a Turing
machine~\cite{bohm64family}, have rather limited set of instructions. Thus we
use more sophisticated methods like Boolean circuits and programming languages
based on QRAM model.

\subsection{Turing machine}
The model of a Turing machine is widely used in classical and quantum complexity
theory. Despite its simplicity it captures the notion of computability.

In what follows by \emph{alphabet} $A=\{a_1,\ldots,a_n\}$ we mean any finite set
of characters or digits. Elements of $A$ are called letters. Set $A^k$ contains
all strings of length $k$ composed from elements of $A$. Elements of $A^k$ are
called \emph{words} and the length of the word $w$ is denoted by $|w|$. The set
of all words over $A$ is denoted by $A^\ast$. Symbol $\epsilon$ is used to
denote an empty word. The complement of language $L\subset A^\ast$ is denoted by
$\bar{L}$ and it is the language defined as $\bar{L}=A^\ast - L$.

\subsubsection{Classical Turing machine}
A Turing machine can operate only using one data structure -- the string of
symbols. Despite its simplicity, this model can simulate any algorithm with
inconsequential loss of efficiency \cite{papadimitriou}. A Classical Turing
machine consists of
\begin{itemize}
 \item an infinitely long tape containing symbols from the finite alphabet $A$,
 \item a head, which is able to read symbols from the tape and write them on
the tape,
 \item memory for storing programme for the machine.
\end{itemize}

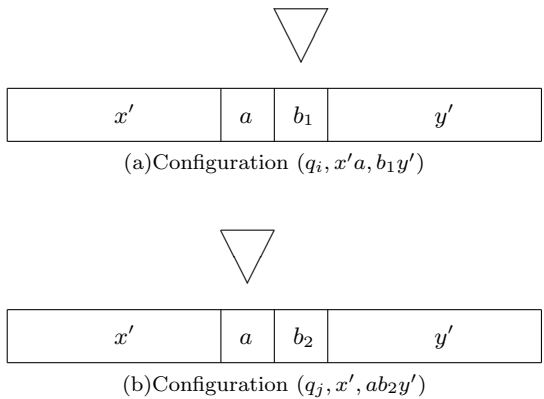
\begin{figure}[ht]
\centering
\subfigure[Configuration $(q_i,x^\prime a,b_1 y^\prime)$]{
 \begin{picture}(200,60)(0,100)
 
 \put(0,120){\line(1,0){200}}
 \put(0,100){\line(1,0){200}}
 
 \put(0,120){\line(0,-1){20}}
 \put(80,120){\line(0,-1){20}}
 \put(100,120){\line(0,-1){20}}
 \put(120,120){\line(0,-1){20}}
 \put(200,120){\line(0,-1){20}}
 
 \put(40,107){$x^\prime$}
 \put(87,107){$a$}
 \put(107,107){$b_1$}
 \put(160,107){$y^\prime$}
 
 \put(100,150){\line(1,0){20}}
 \put(120,150){\line(-1,-2){10}}
 \put(100,150){\line(1,-2){10}}
 
 \end{picture}
 }
 \subfigure[][Configuration $(q_j,x^\prime ,ab_2 y^\prime)$]{
 
  \begin{picture}(200,60)(0,100)
 
 \put(0,120){\line(1,0){200}}
 \put(0,100){\line(1,0){200}}
 
 \put(0,120){\line(0,-1){20}}
 \put(80,120){\line(0,-1){20}}
 \put(100,120){\line(0,-1){20}}
 \put(120,120){\line(0,-1){20}}
 \put(200,120){\line(0,-1){20}}
 
 \put(40,107){$x^\prime$}
 \put(87,107){$a$}
 \put(107,107){$b_2$}
 \put(160,107){$y^\prime$}
 
 \put(80,150){\line(1,0){20}}
 \put(100,150){\line(-1,-2){10}}
 \put(80,150){\line(1,-2){10}}
 
 \end{picture}
 }
 \caption{Computational step of the Turing machine. Configuration $(q_i,x^\prime
a,b_1 y^\prime)$ is presented in (a). If the transition function is defined such
that $\delta(q_i,b_1)=(q_2,b_2,-1)$ this computational step leads to
configuration $(q_j,x^\prime ,ab_2 y^\prime)$ (see (b)).}
 \label{fig:turing-computation}
\end{figure}

The programme for a Turing machine is given in terms of transition function
$\delta$. The schematic illustration of a Turing machine is presented in
Figure~\ref{fig:turing-computation}.

Formally, the classical deterministic Turing machine is defined as follows.
\begin{definition}[Deterministic Turing machine]
 A deterministic Turing machine $M$ over an alphabet $A$ is a
sixtuple $(Q,A,\delta,q_0,q_a,q_r)$, where
\begin{itemize}
 \item $Q$ is the set of internal control states,
 \item $q_0,q_a,q_r\in Q$ are initial, accepting and rejecting states,
 \item $\delta:Q\times A \mapsto Q\times A\times \{-1,0,1\}$ is a transition
function \ie\ the programme of a machine.
\end{itemize}
\end{definition}

By a configuration of machine $M$ we understand a triple $(q_i,x,y)$, $q_i\in
Q$, $x,y\in A^\ast$. This describes a situation where the machine is in the
state $q_i$, the tape contains the word $xy$ and the machine starts to scan the
word $y$. If $x=x^\prime$ and $y=b_1y^\prime$ we can illustrate this situation
as in Figure~\ref{fig:turing-computation}.

The transition from the configuration $c_1$ to the configuration $c_2$ is called
a computational step. We write $c\vdash c'$ if $\delta$ defines the transition
from $c$ to $c'$. In this case $c'$ is called the successor of $c$.

A Turing machine can be used to compute values of functions or to decide about
input words. The computation of a machine with input $w\in A^\ast$ is defined as
a sequence of configurations $c_0, c_1, c_2, \ldots$, such that
$c_0=(q_i,\epsilon,w)$ and $c_i\vdash c_{i+1}$. We say that computation halts if
some $c_i$ has no successor or for configuration $c_i$, the state of the machine
is $q_a$ (machine accepts input) or $q_r$ (machine rejects input).

The computational power of the Turing machine has its limits. Let us define two
important classes of languages.

\begin{definition}
A set of words $L\in A^\ast$ is a recursively enumerable language if there
exists a Turing machine accepting input $w$ iff $w\in L$.
\end{definition}

\begin{definition}
 A set of words $L\in A^\ast$ is a recursive language if there exists a Turing
machine $M$ such that
\begin{itemize}
 \item $M$ accepts $w$ iff $w\in L$,
 \item $M$ halts for any input.
\end{itemize}
\end{definition}

The computational power of the Turing machine is limited by the following
theorem.

\begin{theorem}
 There exists a language $H$ which is recursively enumerable but not recursive.
\end{theorem}

Language $H$ used in the above theorem is defined in halting problem
\cite{papadimitriou}. It consists of all words composed of words encoding Turing
machines and input words for these machines, such that a particular machine
halts on a given word. A universal Turing machine can simulate any machine, thus
for a given input word encoding machine and input for this machine we can easily
perform the required computation.

A deterministic Turing machine is used to measure time complexity of algorithms.
Note that if for some language there exists a Turing machine accepting it, we
can use this machine as an algorithm for solving this problem. Thus we can
measure the running time of the algorithm by counting the number of
computational steps required for Turing machine to output the result.

The time complexity of algorithms can be described using the following
definition.
\begin{definition}
Complexity class $\complexity{TIME}(f(n))$ consists of all languages $L$ such
that there exists a deterministic Turing machine running in time $f(n)$
accepting input $w$ iff $w\in L$.
\end{definition}

In particular complexity class \complexity{P} defined as
\begin{equation}
 \complexity{P} = \bigcup_k \complexity{TIME}(n^k),
\end{equation}
captures the intuitive class of problems which can be solved \emph{easily} on a
Turing machine.

\subsubsection{Nondeterministic and probabilistic computation}
Since one of the main features of quantum computers is their ability to operate
on the superposition of states we can easily extend the classical model of a
probabilistic Turing machine and use it to describe quantum computation. Since
in general many results in the area of algorithms complexity are stated in the
terms of a nondeterministic Turing machine we start by introducing this model.

\begin{definition}[Nondeterministic Turing machine]
A nondeterministic Turing machine $M$ over an alphabet $A$ is a sixtuple
$(Q,A,\delta,q_0,q_a,q_r)$, where
\begin{itemize}
 \item $Q$ is the set of internal control states,
 \item $q_0,q_a,q_r\in Q$ are initial, accepting and rejecting states,
 \item $\delta \subset Q\times A \times Q\times A\times \{-1,0,1\} $ is a
relation.
\end{itemize}
\end{definition}

The last condition in the definition of a nondeterministic machine is the reason
for its power. It also requires to change the definition of acceptance by the
machine.

We say that a nondeterministic Turing machine accepts input $w$ if, for some
initial configuration $(q_i,\epsilon, w)$, computation leads to configuration
$(q_a,a_1,a_2)$ for some words $a_1$ and $a_2$. Thus a nondeterministic machine
accepts the input if there exists some computational path defined by transition
relation $\delta$ leading to an accepting state $q_a$.

The model of a nondeterministic Turing machine is used to define complexity
classes \complexity{NTIME}.

\begin{definition}
Complexity class $\complexity{NTIME}(f(n))$ consists of all languages $L$ such
that there exists a nondeterministic Turing machine running in time $f(n)$
accepting input $w$ iff $w\in L$.
\end{definition}
The most prominent example of these complexity classes is \complexity{NP}, which
is the union of all $\complexity{NTIME}(n^k)$, \ie
\begin{equation}
 \complexity{NP} = \bigcup_k \complexity{NTIME}(n^k).
\end{equation}

\begin{figure}
    \begin{center}
    \includegraphics[viewport=0 0 773 456,width=\columnwidth]{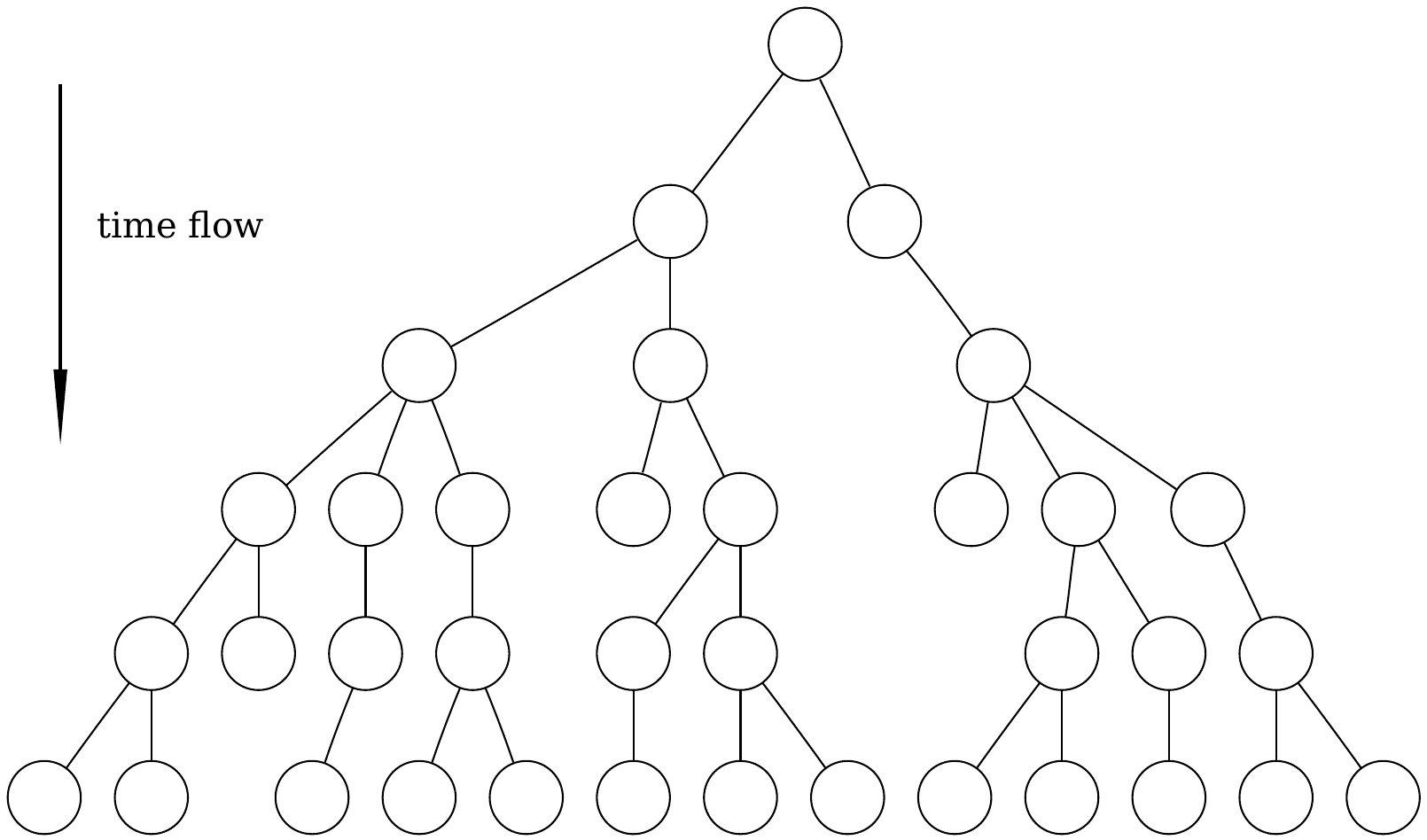}
    \end{center}
    \caption{Schematic illustration of the computational paths of a
    nondeterministic Turing machine~\protect\cite{papadimitriou}. Each circle
    represents the configuration of the machine. The machine can be in many
    configurations simultaneously.} \label{fig:nondeterministic}
\end{figure}

A nondeterministic Turing machine is used as a theoretical model in complexity
theory. However, it is hard to imagine how such device operates. One can
illustrate the computational path of a nondeterministic machine as in
Figure~\ref{fig:nondeterministic}~\cite{papadimitriou}.

Since our aim is to provide the model of a physical device we restrict ourselves
to more realistic model. We can do that by assigning to each element of relation
a number representing probability. In this case we obtain the model of a
probabilistic Turing machine.

\begin{definition}[Probabilistic Turing machine]
 A probabilistic Turing machine $M$ over an alphabet $A$ is a
sixtuple $(Q,A,\delta,q_0,q_a,q_r)$, where
\begin{itemize}
 \item $Q$ is the set of internal control states,
 \item $q_0,q_a,q_r\in Q$ are initial, accepting and rejecting states,
 \item $\delta:Q\times A \times Q\times A\times \{-1,0,1\} \mapsto [0,1] $ is a
transition
probability function \ie
\begin{equation}
 \sum_{(q_2,a_2,d)\in Q\times A\times \{-1,0,1\}} \delta(q_1,a_1,q_2,a_2,d)=1.
\end{equation}
\end{itemize}
\end{definition}

For a moment we can assume that the probabilities of transition used by a
probabilistic Turing machine can be represented only by rational numbers. We do
this to avoid problems with machines operating on arbitrary real numbers. We
will address this problem when extending the above definition to the quantum
case.

The time complexity of computation can be measured in terms of the number of
computational steps of the Turing machine required to execute a programme. Among
important complexity classes we have chosen to point out:
\begin{itemize}
 \item \complexity{P} -- the class of languages for which there exists a
 deterministic Turing machine running in polynomial time,
 \item \complexity{NP} -- the class of languages for which there exists a
 nondeterministic Turing machine running in polynomial time,
 \item \complexity{RP} -- the class of languages $L$ for which there exists a
 probabilistic Turing machine $M$ such that: $M$ accepts input $w$ with
 probability at least $\frac{1}{2}$ if $w\in L$ and always rejects $w$ if
 $w\not\in L$,
 \item \complexity{coRP} -- the class of languages $L$ for which $\bar{L}$ is in
 \complexity{RP},
 \item \complexity{ZPP} -- $\complexity{RP}\cap\complexity{coRP}$.
\end{itemize}

More examples of interesting complexity classes and computational problems
related to them can be found in~\cite{complexityzoo}.
\subsubsection{Quantum Turing machine}
A quantum Turing machine was introduced by Deutsch \cite{deutsch85principle}.
This model is equivalent to a quantum circuit model
\cite{yao93complexity,nishimura02computational}. However, it is very
inconvenient for describing quantum algorithms since the state of a head and the
state of a tape are described by state vectors.

A quantum Turing machine consists of
\begin{itemize}
\item Processor: $M$ 2-state observables $\set{n_i}{i\in_{\Z_M}}$.
\item Memory: infinite sequence of 2-state observables $\set{m_i}{i\in\Z}$.
\item Observable $x$, which represents the address of the current head position.
\end{itemize}
The state of the machine is described by the vector $\ket{\psi(t)} =
\ket{x;n_0,n_1,\ldots;m}$ in the Hilbert space $\HS{H}$ associated with the
machine.

At the moment $t=0$ the state of the machine is described by the vectors
$\ket{\psi(0)} = \sum_{m}a_m\ket{0;0,\ldots,0;\ldots,0,0,0,\ldots}$ 
such that 
\begin{equation}
  \sum_{i}|a_i|^2=1.
\end{equation}

The evolution of the quantum Turing machine is described by the unitary operator
$U$ acting on $\HS{H}$.

A classical probabilistic (or nondeterministic) Turing machine can be described
as a quantum Turing machine such that, at each step of its evolution, the state
of the machine is represented by the base vector.

The formal definition of the quantum Turing machine was introduced in
\cite{bernstein97complexity}.

It is common to use real numbers as amplitudes when describing the state of
quantum systems during quantum computation. To avoid problems with an arbitrary
real number we introduce the class of numbers which can be used as amplitudes
for amplitude transition functions of the quantum Turing machine.

Let us denote by $\widetilde{\Cplx}$ the set of complex numbers $c\in\Cplx$,
such that there exists a deterministic Turing machine, which allows to calculate
$\re{c}$ and $\im{c}$ with accuracy $\frac{1}{2^n}$ in time polynomial in $n$.

\begin{definition}[Quantum Turing Machine]
A quantum Turing machine (QTM) $M$ over an alphabet $A$ is a sixtuple
$(Q,A,\delta,q_0,q_a,q_r)$, where
\begin{itemize}
    \item $Q$ is the set of internal control states,
    \item $q_0,q_a,q_r\in Q$ are initial, accepting and rejecting states,
    \item $\delta:Q\times A \times Q\times A\times \{-1,0,1\} \mapsto
    \widetilde{\Cplx} $ is a transition amplitude function \ie
\begin{equation}
    \sum_{(q_2,a_2,d)\in Q\times A\times \{-1,0,1\}}
    |\delta(q_1,a_1,q_2,a_2,d)|^2=1.
\end{equation}
\end{itemize}
\end{definition}

Reversible classical Turing machines (\ie Turing machines with reversible
transition function) can be viewed as particular examples of quantum machines.
Since any classical algorithm can be transformed into reversible form, it is
possible to simulate a classical Turing machine using quantum Turing machine.

\subsubsection{Quantum complexity}\label{sec:quantum-complexity}
Quantum Turing machine allows for rigorous analysis of algorithms. This is
important since the main goal of quantum information theory is to provide some
gain in terms of speed or memory with respect to classical algorithms. It should
be stressed that at the moment no formal proof has been given that a quantum
Turing machine is more powerful than a classical Turing
machine~\cite{fortnow03one}.

In this section we give some results concerning quantum complexity theory. See
also \cite{bernstein97complexity,vaziriani05survey} for a introduction to this
subject.

In analogy to classical case it is possible to define complexity classes for the
quantum Turing machine. The most important complexity class is this case
is~$\complexity{BQP}$.
 
\begin{definition}
Complexity class $\complexity{BQP}$ contains languages $L$ for which there
exists a quantum Turing machine running in polynomial time such that, for any
input word $x$, this word is accepted with probability at least $\frac{3}{4}$ if
$x\in L$ and is rejected with probability at least $\frac{3}{4}$ if $x\not\in
L$.
\end{definition}

Class \complexity{BQP} is a quantum counterpart of the classical class
\complexity{BPP}.

\begin{definition}
Complexity class $\complexity{BPP}$ contains languages $L$ for which there
exists a nondeterministic Turing machine running in polynomial time such that,
for any input word $x$, this word is accepted with probability at least
$\frac{3}{4}$ if $x\in L$ and is rejected with probability at least
$\frac{3}{4}$ if $x\not\in L$.
\end{definition}

Since many results in complexity theory are stated in terms of oracles, we
define an oracle as follows.

\begin{definition}\label{def:oracle}
An oracle or black box is an imaginary machine which can decide certain problems
in a single operation.
\end{definition}

We use notation $\complexity{A}^\complexity{B}$ to describe the class of
problems solvable by an algorithm in class \complexity{A} with an oracle
for the language \complexity{B}.

It was shown \cite{bernstein97complexity} that the quantum complexity classes
are related as follows. 

\begin{theorem} Complexity classes fulfil the following inequality
\begin{equation}
\complexity{BPP}\subseteq\complexity{BQP}\subseteq
\complexity{P}^\complexity{\#P}.
\end{equation}
\end{theorem}

Complexity class $\complexity{\#P}$ consists of problems of the form
\emph{compute f(x), where f is the number of accepting paths of an
$\complexity{NP}$ machine}. For example problem \complexity{\#SAT} formulated
below is in \complexity{\#P}.

\begin{problem}[\complexity{\#SAT}]
For a given Boolean formula, compute how many satisfying true assignments it
has.
\end{problem}

Complexity class $\complexity{P}^\complexity{\#P}$ consists of all problems
solvable by a machine running in polynomial time which can use oracle for
solving problems in~\complexity{\#P}.

Complexity ZOO \cite{complexityzoo} contains the description of complexity
classes and many famous problems from complexity theory. The complete
introduction to the complexity theory can be found in \cite{papadimitriou}.
Theory of \complexity{NP}-completeness with many examples of problems from this
class is presented in~\cite{np}.

Many important results and basic definitions concerning quantum complexity
theory can be found in \cite{bernstein97complexity}. The proof of equivalence
between quantum circuit and quantum Turing machine was given in
\cite{yao93complexity}. An interesting discussion of quantum complexity classes
and relation of \complexity{BQP} class to classical classes can be found
in~\cite{fortnow03one}.

\subsection{Quantum computational networks}
After presenting the basic facts about Turing machines we are ready to introduce
more usable models of computing devices. We start by defining Boolean circuits
and extending this model to the quantum case. 

\subsubsection{Boolean circuits}
Boolean circuits are used to compute functions of the form
\begin{equation}
 f:\{0,1\}^m\mapsto\{0,1\}^n.
\end{equation}
Basic gates (functions) which can be used to define such circuits are 
\begin{itemize}
 \item $\wedge:\{0,1\}^2\mapsto\{0,1\}$, $\wedge(x,y)=1 \Leftrightarrow x=y=1$
(\emph{logical and}),
 \item $\vee:\{0,1\}^2\mapsto\{0,1\}$, $\vee(x,y)=0 \Leftrightarrow x=y=0$
(\emph{logical or}),
 \item $\sim:\{0,1\}\mapsto\{0,1\}$,  $\sim(x)=1-x$ (\emph{logical not}).
\end{itemize}

The set of gates is called universal if all functions $\{0,1\}^n\mapsto \{0,1\}$
can be constructed using the gates from this set. It is easy to show that the
set of functions composed of $\sim$, $\vee$ and $\wedge$ is universal. Thus it
is possible to compute any functions $\{0,1\}^n\mapsto \{0,1\}^m$ using only
these functions. The full characteristic of universal sets of functions was
given by Post in 1949~\cite{zwick95notes}.

Using the above set of functions a Boolean circuit is defined as follows.
\begin{definition}[Boolean circuit]
A Boolean circuit is an acyclic direct graph with nodes labelled by input
variables, output variables or logical gates $\vee$, $\wedge$ or $\sim$.
\end{definition}
Input variable node has no incoming arrow while output variable node has no
outcoming arrows. The example of a Boolean circuit computing the sum of bits
$x_1$ and $x_2$ is given in Figure~\ref{fig:bool-circ-ex}.

\begin{figure}[ht]
\begin{center}
 \begin{picture}(300,82)(0,0)
 \setlength{\unitlength}{0.9\unitlength}
 \put (20,10){\circle{20}}
 \put (15.5,8){$x_1$}
 
 \put(30,10){\vector(1,0){80}}
 
 \put (120,10){\circle{20}}
 \put (116.5,7){$\vee$}
 
 \put(130,10){\vector(1,0){30}}

 \put (170,10){\circle{20}}
 \put (166.5,7){$\wedge$}

 \put(180,10){\vector(1,0){30}}

 \put (220,10){\circle{20}}
 \put (216,8){$y_1$}
 
 
 \put (20,90){\circle{20}}
 \put (15.5,88){$x_2$}
 
  \put(30,90){\vector(1,0){30}}
 
 \put (70,90){\circle{20}}
 \put (65.5,87.5){$\sim$}
 
 \put(80,90){\vector(1,0){30}}
 
 \put (120,90){\circle{20}}
 \put (116,87){$\vee$}

 \put(130,90){\vector(1,0){30}}

 \put (170,90){\circle{20}}
 \put (165.5,87){$\sim$}

 \put(180,90){\vector(1,0){30}}

 \put (220,90){\circle{20}}
 \put (215.5,88){$y_2$}
 
 
 \put (91,63){\circle{20}}
 \put (86.5,60.5){$\sim$}

 \put (29,15){\vector(4,3){54.5}}
 \put (98,69){\vector(4,3){16.5}}
 
 \put (29,85){\vector(4,-3){87.5}}
 
 \end{picture}
\end{center}
\caption{The example of a Boolean circuit
computing the sum of bits $x_1$ and $x_2$ \protect\cite{hirvensalo}. Nodes labelled
$x_1$ and $x_2$ represent input variables and nodes labelled $y_1$ and $y_2$
represent output variables.}
\label{fig:bool-circ-ex}
\end{figure}

Note that in general it is possible to define a Boolean circuit using different
sets of elementary functions. Since functions $\vee$, $\wedge$ and $\sim$
provide a universal set of gates we defined Boolean circuit using these
particular functions.

Function $f:\{0,1\}^m\mapsto \{0,1\}$ is defined on the binary string of
arbitrary length. Let $f_n:\{0,1\}^m\mapsto \{0,1\}^n$ be a restriction of $f$
to $\{0,1\}^n$. For each such restriction there is a Boolean circuit $C_n$
computing $f_n$. We say that $C_0, C_1, C_2, \ldots$ is a \emph{family of
Boolean circuits} computing $f$.

Note that any binary language $L\subset \{0,1\}^\ast$ can be accepted by some
family of circuits. But since we need to know the value of $f_n$ to construct a
circuit $C_n$ such family is not an algorithmic device at all. We can state that
there exists a family accepting the language, but we do not know how to build it
\cite{papadimitriou}.

To show how Boolean circuits are related to Turing machines we introduce
uniformly generated circuits.

\begin{definition}
 We say that language $L\in A^\ast$ has uniformly polynomial circuits if there
exists a Turing machine $M$ that an input $\underbrace{1\ldots 1}_n$ outputs
the graph of circuit $C_n$ using space $\BigOh{\log n}$, and the family
$C_0,C_1, \ldots$ accepts $L$.
\end{definition}

The following theorem provides a link between uniformly generated circuits and
Turing machines.

\begin{theorem}
A language $L$ has uniformly polynomial circuit iff $L\in\complexity{P}$.
\end{theorem}

Quantum circuits model is an analogous to uniformly polynomial circuits. They
can be introduced as the straightforward generalisation of reversible circuits.

\subsubsection{Reversible circuits}
The evolution of isolated quantum systems is described by a unitary operator
$U$. The main difference with respect to classical evolution is that this type
of evolution is \emph{reversible}.

Before introducing a quantum circuit we define a reversible Boolean circuit
\begin{definition}[Reversible gate]
A classical reversible function (gate) $\{0,1\}^m\mapsto\{0,1\}^m$ is a
permutation.
\end{definition}

\begin{definition}
 A reversible Boolean circuit is a Boolean circuit composed of reversible gates.
\end{definition}

The important fact expressed by the following theorem allows us to simulate any
classical computation on a quantum machine described using a reversible circuit
\begin{theorem}
All Boolean circuits can be simulated using reversible Boolean circuits.
\end{theorem}

Like in the case of nonreversible circuit one can introduce the universal set of
functions for reversible circuits. 

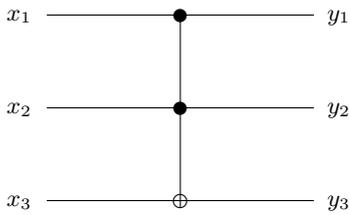
\begin{figure}[ht]
 \centering
 \begin{picture}(100,80)(0,0)
  \put(50,70){\circle*{5}}
  \put(50,35){\circle*{5}}
  \put(50,0){\circle{5}}
  
  \put(0,0){\line(1,0){100}}
  \put(0,35){\line(1,0){100}}
  \put(0,70){\line(1,0){100}}
  
  \put(50,70){\line(0,-1){72.5}}
  
  \put(-15,-2){$x_3$}
  \put(-15,33){$x_2$}
  \put(-15,68){$x_1$}
  
  \put(105,-2){$y_3$}
  \put(105,33){$y_2$}
  \put(105,68){$y_1$}
 \end{picture}

 \caption{Classical Toffoli gate is universal for reversible
 circuits. It was also used to provide the universal
 set of quantum gates \protect\cite{deutsch89networks}.}
 \label{fig:toffoli}
\end{figure}

The important example of a gate universal for reversible Boolean circuits is a
Toffoli gate. The graphical representation of this gate is presented in
Figure~\ref{fig:toffoli}. The following theorem was proved by Toffoli
\cite{toffoli81bicontinuous}.

\begin{theorem}
A Toffoli gate is a universal reversible gate.
\end{theorem}

As we will see in the following section it is possible to introduce two-bit
quantum gates which are universal for quantum circuits. This is impossible in
the classical case and one needs at least a three-bit gate to construct the
universal set of reversible gates.

In particular, any reversible circuit is automatically a quantum circuit.
However, quantum circuits offer much more diversity in terms of the number of
allowed operations.

\subsubsection{Quantum circuits}
The computational process of the quantum Turing machine is complicated since
data as well as control variables can be in a superposition of base states. To
provide more convenient method of describing quantum algorithms one can use a
quantum circuits model. This model is sometimes called a \emph{quantum gate
array} model.

Quantum circuits model was first introduced by Deutsch in
\cite{deutsch89networks} and it is the most commonly used notation for quantum
algorithms. It is much easier to imagine than the quantum Turing machine since
the control variables (executed steps and their number) are classical. There are
only quantum data (\eg\ qubits or qudits and unitary gates) in a quantum
circuit.

A quantum circuit consists of the following elements (see Table~\ref{tab:gates}):
\begin{itemize}
  \item the finite sequence of \emph{wires} representing qubits or sequences of
  qubits (quantum registers),
  \item quantum gates representing elementary operations from the particular set
  of operations implemented on a quantum machine,
  \item measurement gates representing a measurement operation, which is usually
  executed as the final step of a quantum algorithm. It is commonly assumed that
  it is possible to perform the measurement on each qubit in canonical basis $\{
  \ket{0}, \ket{1} \}$ which corresponds to the measurement of the $S_z$
  observable.
\end{itemize} 

The concept of a quantum circuit is the natural generalisation of acyclic logic
circuits studied in classical computer science. Quantum gates have the same
number of inputs as outputs. Each $n$-qubit quantum gate represents the
$2^n$-dimensional unitary operation of the group $SU(2^n)$, \ie\ generalised
rotation in a complex Hilbert space.

The main advantage of this model is its simplicity. It also provides very
convenient representation of physical evolution in quantum systems.

From the mathematical point of view quantum gates are unitary matrices acting on
$n$-dimensional Hilbert space. They represent the evolution of an isolated quantum
system \cite{nielsenchuang}.

The problem of constructing new quantum algorithms requires more careful study
of operations used in quantum circuit model. In particular we are interested in
efficient decomposition of quantum gates into elementary operations.

We start by providing basic charactersistic of unitary matrices
\cite{barenco95elementary,nielsenchuang}
\begin{theorem}\label{the:unitary-decomposition}
Every unitary $2\times 2$ matrix $G\in U(2)$ can be decomposed using elementary
rotations as
\begin{equation}
G = \Phi(\delta) R_z(\alpha) R_y(\theta) R_z(\beta)
\end{equation}
where
$$
  \Phi(\xi)=
    \left(
      \begin{array}{cc}
      e^{i\xi} & 0\\
      0 & e^{i\xi}
      \end{array}
    \right),
$$
$$
  R_y(\xi)=
    \left(
        \begin{array}{cc}
        \cos({\xi}/{2}) & \sin({\xi}/{2})\\
        -\sin({\xi}/{2}) & \cos({\xi}/{2})
        \end{array}
    \right),
$$
and
$$
  R_z(\xi)=
    \left(
        \begin{array}{cc}
        e^{i\frac{\xi}{2}} & 0\\
        0 & e^{-i\frac{\xi}{2}}
        \end{array}
    \right).
$$
\end{theorem}

We introduce the definition of quantum gates as stated in \cite{hirvensalo}.
\begin{definition}\label{def:quantum-gate}
A quantum gate $U$ acting on $m$ qubits is a unitary mapping on
$\Cplx^{2^m}\equiv\underbrace{\Cplx^2\otimes\ldots\otimes\Cplx^2}_{m\
\mathrm{times}}$
\begin{equation}
 U: \Cplx^{2^m} \mapsto\Cplx^{2^m} ,
\end{equation}
which operates on the fixed number of qubits.
\end{definition}

Formally, a quantum circuit is defined as the unitary mapping which can be
decomposed into the sequence of elementary gates.
\begin{definition}
A quantum circuit on $m$ qubits is a unitary mapping on $\Cplx^{2^m}$, which can
be represented as a concatenation of a finite set of quantum gates.
\end{definition}

Any reversible classical gate is also a quantum gate. In particular logical gate
$\sim$ (negation) is represented by quantum gate $NOT$, which is realized by
$\sigma_x$ Pauli matrix.

As we know any Boolean circuit can be simulated by a reversible circuit and thus
any function computed by a Boolean circuit can be computed using a quantum
circuit. Since a quantum circuit operates on a vector in complex Hilbert space
it allows for new operations typical for this model.

The first example of quantum gate which has no classical counterpart is
$\sqrt{NOT}$ gate. It has the following property
\begin{equation}
 \sqrt{NOT}\sqrt{NOT}=NOT,
\end{equation}
which cannot be fulfilled by any classical Boolean function
$\{0,1\}\mapsto\{0,1\}$. 
Gate $\sqrt{N}$ is represented by the unitary matrix
\begin{equation}
 \sqrt{NOT}=\frac{1}{2}\left(\begin{array}{cc}
             1+i & 1-i \\
             1-i & 1+i
            \end{array}\right).
\end{equation}

Another example is Hadamard gate $H$. This gate is used to introduce the
superposition of base states. It acts on the base state as
\begin{eqnarray}
 H\ket{0} = \frac{1}{\sqrt{2}}\left(\ket{0}+\ket{1}\right),
 H\ket{1} = \frac{1}{\sqrt{2}}\left(\ket{0}-\ket{1}\right).
\end{eqnarray}

If the gate $G$ is a quantum gate acting on one qubit it is possible to
construct the family of operators acting on many qubits. The particularly
important class of multiqubit operations is the class of controlled operations.
\begin{definition}[Controlled gate]\label{def:one-controlled}
Let $G$ be a $2\times 2$ unitary matrix representing a quantum gate. Operator
\begin{equation}
\Proj{1}\otimes G + \Proj{0}\otimes \Id
\end{equation}
acting on two qubits, is called a controlled-$G$ gate.
\end{definition}
Here $A\otimes B$ denotes the tensor product of gates (unitary operator) $A$ and
$B$, and $\Id$ is an identity matrix. If in the above definition we take $G=NOT$
we get 
\begin{equation}
 \Proj{1}\otimes \sigma_x+\Proj{0}\otimes \Id = \left(\begin{array}{cccc}
                                                 1 & 0 & 0 & 0\\
                                                 0 & 1 & 0 & 0\\
                                                 0 & 0 & 0 & 1\\
                                                 0 & 0 & 1 & 0
                                                \end{array}\right),
\end{equation}
which is the definition of $CNOT$(controlled-$NOT$) gate. This gate can be used
to construct the universal set of quantum gates. This gate also allows to
introduce entangled states during computation
\begin{eqnarray*}
 CNOT (H\otimes \Id)\ket{00} &=&CNOT\frac{1}{\sqrt{2}}\left(\ket{00}+\ket{10}\right)\\
 &=&\frac{1}{\sqrt{2 } } \left(\ket{00}+\ket{11}\right)
\end{eqnarray*}
The classical counterpart of $CNOT$ gate is $XOR$ gate.

\begin{table}[hb]
\begin{center}
\begin{tabular}{|c|c|c|}
\hline
$x_1$ & $x_2$ & $x_1$ XOR $x_2$  \\ \hline
0 & 0 & 0 \\ 
0 & 1 & 1 \\ 
1 & 0 & 1 \\ 
1 & 1 & 0 \\ \hline
\end{tabular}
\caption{Logical values for XOR gate. Quantum 
CNOT gate computes value of $x_1$ XOR $x_2$ in the first register and stores
values of $x_2$ in the second register.}
\end{center}
 \label{tab:xor}
\end{table}

\begin{figure}[ht]
  \begin{center}
	\includegraphics[scale=1.2]{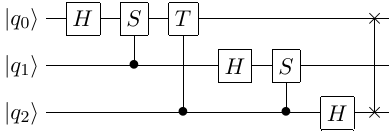}
  \end{center}
  \caption{Quantum circuit
  representing quantum Fourier transform for three qubits. Elementary gates used
  in this circuit are described in Table~\ref{tab:gates}.}
  \label{fig:qft3}
\end{figure}

Other examples of single-qubit and two-qubit quantum gates are presented in
Table~\ref{tab:gates}. In Figure~\ref{fig:qft3} a quantum circuit for quantum
Fourier transform on three qubits is presented.

\begin{table*}[htp!]
  \centering
  \begin{tabular*}{\textwidth}{@{\extracolsep{\fill}}ccc}
    The name of the gate & Graphical representation & Mathematical form \\
    \hline\\
    Hadamard
    &
    \Qcircuit @C=1.2em @R=1.4em {
    & \gate{H} & \qw
    }     
    & 
    $
    \frac{1}{\sqrt{2}}
    \left(
    \begin{array}{cc}
      1 & 1 \\
      1 & -1 \\
    \end{array} 
    \right)
    $ \\[2em]
    Pauli X
    &
    \Qcircuit @C=1.2em @R=1.4em {
    & \gate{X} & \qw
    }     
    & 
    $
    \left(
    \begin{array}{cc}
      0 & 1 \\
      1 & 0 \\
    \end{array} 
    \right)
    $ \\[2em]
    Pauli Y
    &
    \Qcircuit @C=1.2em @R=1.4em {
    & \gate{Y} & \qw
    }     
    & 
    $
    \left(
    \begin{array}{cc}
      0 & -i \\
      i & 0 \\
    \end{array} 
    \right)
    $ \\[2em]
    Pauli Z
    &
    \Qcircuit @C=1.2em @R=1.4em {
    & \gate{Z} & \qw
    }     
    & 
    $
    \left(
    \begin{array}{cc}
      1 & 0 \\
      0 & -1 \\
    \end{array} 
    \right)
    $ \\[2em]
    Phase
    &
    \Qcircuit @C=1.2em @R=1.4em {
    & \gate{S} & \qw
    }     
    & 
    $
    \left(
    \begin{array}{cc}
      1 & 0 \\
      0 & i \\
    \end{array} 
    \right)
    $ \\[2em]
    $\pi/8$
    &
    \Qcircuit @C=1.2em @R=1.4em {
    & \gate{T} & \qw
    }     
    & 
    $
    \left(
    \begin{array}{cc}
      1 & 0 \\
      0 & e^{i\pi/4} \\
    \end{array} 
    \right)
    $ \\[2em]
    CNOT
    &
    \Qcircuit @C=1.2em @R=1.4em {
    & \ctrl{1} & \qw \\
    & \targ & \qw
    }
    &  
    $
    \left(
    \begin{array}{cccc}
      1 & 0 & 0 & 0 \\
      0 & 1 & 0 & 0 \\
      0 & 0 & 0 & 1 \\
      0 & 0 & 1 & 0 
    \end{array} 
    \right)
    $ \\[3em]
    SWAP
    &
    \Qcircuit @C=1.2em @R=1.4em {
    & \qswap & \qw \\
    & \qswap \qwx & \qw 
    }
    &  
    $
    \left(
    \begin{array}{cccc}
      1 & 0 & 0 & 0 \\
      0 & 0 & 1 & 0 \\
      0 & 1 & 0 & 0 \\
      0 & 0 & 0 & 1 
    \end{array} 
    \right)
    $ \\[2.4em] 
    Measurement
    &
    \Qcircuit @C=1.2em @R=1.4em {
    & \meter & \cw
    }     
    & 
    $
    \left\{
    \left(
    \begin{array}{cc}
      1 & 0 \\
      0 & 0 \\
    \end{array}
    \right)
    ,  
    \left(
    \begin{array}{cc}
      0 & 0 \\
      0 & 1 \\
    \end{array}
    \right)
    \right\}
    $ \\[1.4em]
    qubit
    &
    \Qcircuit @C=1.2em @R=1.4em {
    & \qw & \qw
    }     
    & 
    wire $\equiv$ single qubit
    \\[1.3em]
    $n$ qubits
    &
    \Qcircuit @C=1.2em @R=1.2em {
    & {/} \qw &  \qw
    }     
    & 
    wire representing $n$ qubits
    \\[1.3em]
    classical bit
    &
    \Qcircuit @C=1.2em @R=1.2em {
    & \cw & \cw
    }     
    & 
    double wire $\equiv$ single bit
    \\[.8em]
  \end{tabular*}
    \caption{Basic gates used in quantum
    circuits with their graphical representation and mathematical form. Note
    that measurement gate is represented in Kraus form, since it is the example
    of non-unitary quantum evolution.}
\label{tab:gates}
\end{table*}

\begin{figure}[ht]\index{quantum!circuit}
  \begin{center}
	\includegraphics[scale=1.2]{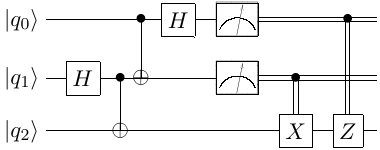}
  \end{center}
  \caption{Circuit for quantum teleportation.
  Double lines represent the operation which is executed depending on the
  classical data obtained after the measurement on a subsystem.}
  \label{fig:teleport}
\end{figure}

One can extend Definition \ref{def:one-controlled} and introduce quantum gates
with many controlled qubits.
\begin{definition}\label{def:m-toffoli}
 Let $G$ be a $2\times 2$ unitary matrix. Quantum gate defined as
 \begin{equation}
  \Proj{\underbrace{1\ldots1}_{n-1}}\otimes G +
\sum_{l\not=\underbrace{1\ldots1}_{n-1}}\Proj{l}\otimes \Id
 \end{equation}
is called $(n-1)$-controlled $G$ gate. We denote this gate by
$\wedge_{n-1}(G)$.
\end{definition}

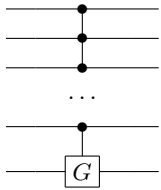
\begin{figure}[ht]
\[ \Qcircuit @C=1.2em @R=1.2em {
   & \qw & \ctrl{1} & \qw & \qw \\
   & \qw & \ctrl{1} & \qw & \qw \\
   & \qw & \ctrl{0} & \qw & \qw \\
   &  & \ldots &  &  \\
   & \qw & \ctrl{1} & \qw & \qw\\
   & \qw & \gate{G} & \qw & \qw
}\]
    \caption{Generalised quantum Toffoli gate
    acting on $n$ qubits. Gate $G$ is controlled by the state of $n-1$ qubits
    according to Definition \ref{def:m-toffoli}.}
    \label{fig:m-toffoli}
\end{figure}

This gate $\wedge_{n-1}(G)$ is sometimes referred to as a generalised Toffoli
gate or a Toffoli gate with $m$ controlled qubits. Graphical representation of
this gate is presented in Figure \ref{fig:m-toffoli}.

The important feature of quantum circuits is expressed by the following
universality property~\cite{barenco95elementary}.
\begin{theorem}
The set of gates consisting of all one-qubit gates $U(2)$ and one two-qubit CNOT
gate is universal in the sense that any $n$-qubit operation can be expressed as
the composition of these gates.
\end{theorem}

Note that, in contrast to the classical case, where one needs at least three-bit
gates to construct a universal set, quantum circuits can be simulated using one
two-qubit universal gate.

In order to implement a quantum algorithm one has to decompose many qubit
quantum gates into elementary gates. It has been shown that almost any $n$-qubit
quantum gate ($n\geq2$) can be used to the build a universal set of gates
\cite{deutsch95universality} in the sense that any unitary operation on the
arbitrary number of qubits can be expressed as the composition of gates from
this set. In fact the set consisting of two-qubit exclusive-or (XOR) quantum
gate and all single-qubit gates is also universal~\cite{barenco95elementary}.

Let us assume that we have the set of gates containing only CNOT and one-qubit
gates. In~\cite{shende04minimal} theoretical lower bound for the number of gates
required to simulate a circuit using these gates was derived. The efficient
method of elementary gates sequence synthesis for an arbitrary unitary gate was
presented in~\cite{mottonen04general}.

\begin{theorem}[Shende-Markov-Bullock]\label{theor:lower-bound-decompose}
Almost all $n$-qubit operators cannot be simulated by a circuit with fewer than
$\lceil \frac{1}{4}[4^n-3n-1] \rceil$ CNOT gates.
\end{theorem}

In \cite{vartiainen04efficient} the construction providing the efficient way of
implementing arbitrary quantum gates was described. The resulting circuit has
complexity $\BigOh{4^n}$ which coincides with lower bound from
Theorem~\ref{theor:lower-bound-decompose}.

It is useful to provide more details about the special case, when one uses gates
with many controlled and one target qubits. The following results were proved in
\cite{barenco95elementary}.

\begin{theorem}\label{theor:decompose-1}
For any single-qubit gate $U$ the gate $\wedge_{n-1}(U)$ can be simulated in
terms of $\Theta(n^2)$ basic operations.
\end{theorem}

In many situations it is useful to construct a circuit which approximates the
required circuit. We say that quantum circuits approximate other circuits with
accuracy $\varepsilon$ if the distance (in terms of Euclidean norm) between
unitary transformations associated with these circuits is at
most~$\varepsilon$~\cite{barenco95elementary}. 

\begin{theorem}
For any single-qubit gate $U$ and $\varepsilon>0$ gate $\wedge_{n-1}(U)$ can be
approximated with accuracy $\varepsilon$ using
$\Theta(n\log\frac{1}{\varepsilon})$ basic operations.
\end{theorem}

Note that the efficient decomposition of a quantum circuit is crucial in
physical implementation of quantum information processing. In particular case
decomposition can be optimised using the set of elementary gates specific for
target architecture. CNOT gates are of big importance since they allow to
introduce entangled states during computation. It is also hard to physically
realise CNOT gate since one needs to control physical interaction between
qubits. 

One should also note that for some classes of quantum circuits it is possible to
construct their classical counterparts, which can be used to simulate quantum
computation performed by these circuits efficiently. The most notable class
having this property is a class of circuits $\mathrm{CHP}$ class, which consists
of stabilizer circuits, \ie\ circuits consisting solely of CNOT, Hadamard and
phase gates~\cite{aaronson04improved}. This property is known as so called
Gottesman-Knill theorem.

\begin{theorem}[Gottesman-Knill]
Any stabilizer circuit can be efficiently simulated on a classical machine.
\end{theorem}

It is worth noting that gates used to construct stabilizer circuits do not
provide an universal set of gates. Nevertheless, such circuits can produce
highly entangled states.

\subsection{Random access machines}
\begin{figure}[ht]
 \centering
 \includegraphics[viewport=0 0 518 254,width=\columnwidth]{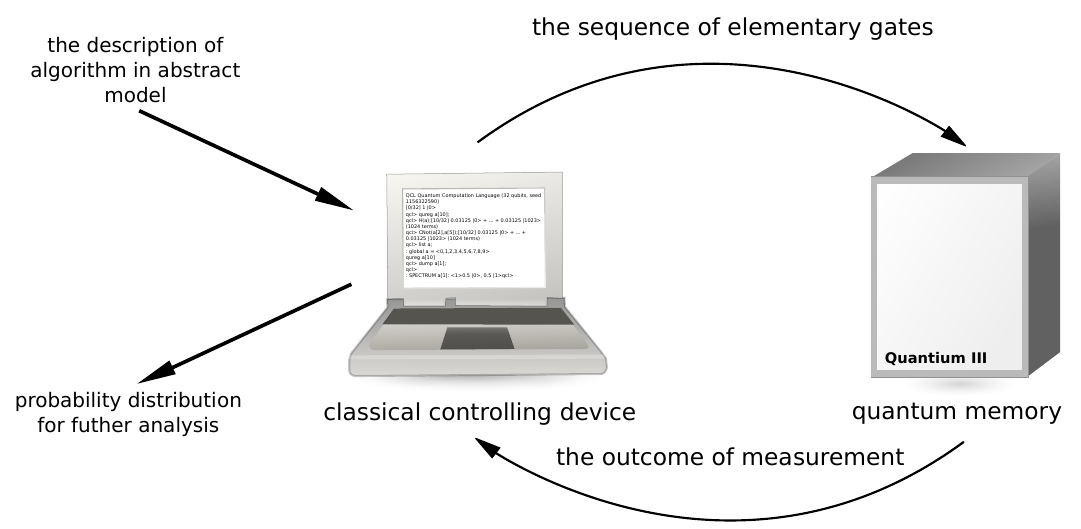}
 \caption{The model of classically controlled
 quantum machine \protect\cite{oemerPHD}. Classical computer is responsible for
 performing unitary operations on quantum memory. The results of quantum
 computation are received in the form of measurement results.}
 \label{fig:qram}
\end{figure}

Quantum circuit model does not provide a mechanism for controlling with
classical machine the operations on quantum memory. Usually quantum algorithms
are described using mathematical representation, quantum circuits and classical
algorithms~\cite{knill96conventions}. The model of quantum random access machine
is built on an assumption that the quantum computer has to be controlled by a
classical device~\cite{oemerPHD}. Schematic presentation of such architecture is
provided in Figure~\ref{fig:qram}.

Quantum random access machine is interesting for us since it provides a
convenient model for developing quantum programming languages. However, these
languages are our main area of interest. We see no point in providing the
detailed description of this model as it is given in \cite{oemerPHD} together
with the description of hybrid architecture used in quantum programming.

\subsubsection{Classical RAM model}
The classical model of random access machine (RAM) is the example of more
general register machines \cite{cook73time,shepherdson63computability}.

The random access machine consists of an unbounded sequence of memory registers
and a finite number of arithmetic registers. Each register may hold an arbitrary
integer number. The programme for the RAM is a finite sequence of instructions
$\Pi=(\pi_1,\ldots,\pi_n)$. At each step of execution register $i$ holds an
integer $r_i$ and the machine executes instruction $\pi_\kappa$, where $\kappa$
is the value of the programme counter. Arithmetic operations are allowed to
compute the address of a memory register.

Despite the difference in the construction between a Turing machine and RAM, it
can be easily shown that a Turing machine can simulate any RAM machine with
polynomial slow-down only \cite{papadimitriou}.

It is worth noting that programming languages can be defined without using RAM
model. Interesting programming language for a Turing machine
$\mathcal{P}^{\prime\prime}$, providing the minimal set of instructions, was
introduced by B\"ohm in~\cite{bohm64family}.

\subsubsection{Quantum RAM model}\label{sec:qram-model}
Quantum random access machine (QRAM) model is the extension of the classical
RAM. QRAM can exploit quantum resources and, at the same time, can be used to
perform any kind of classical computation. It allows us to control operations
performed on quantum registers and provides the set of instructions for defining
them.

Recently a new model of sequential quantum random machine (SQRAM) has been
proposed. Instruction set for this model and compilation of high-level languages
is discussed in \cite{nagarajan07sqram}. However, it is very similar to QRAM
model.

The quantum part of QRAM model is used to generate probability distribution.
This is achieved by performing measurement on quantum registers. The obtained
probability distribution has to be analysed using a classical computer.

\subsubsection{Quantum pseudocode}
Quantum algorithms are, in most of the cases, described using the mixture of
quantum gates, mathematical formulas and classical algorithms. The first attempt
to provide a uniform method of describing quantum algorithms was made in
\cite{cleve96schumacher}, where the author introduced a high-level notation
based on the notation known from computer science textbooks~\cite{cormen}.

In~\cite{knill96conventions} the first formalised language for description of
quantum algorithms was introduced. Moreover, it was tightly connected with the
model of quantum machine called quantum random access machine (QRAM).

Quantum pseudocode proposed by Knill \cite{knill96conventions} is based on
conventions for classical pseudocode proposed in \cite[Chapter 1]{cormen}.
Classical pseudocode was designed to be readable by professional programmers, as
well as people who had done a little programming. Quantum pseudocode introduces
operations on quantum registers. It also allows to distinguish between classical
and quantum registers.

Quantum registers are distinguished by underlining them. They can be introduced
by applying quantum operations to classical registers or by calling a subroutine
which returns a quantum state. In order to convert a quantum register into a
classical register measurement operation has to be perfomed.

The example of quantum pseudocode is presented in Listing
\ref{lst:qpc-example}. It shows the main advantage of QRAM model over quantum
circuits model -- the ability to incorporate classical control into the
description of quantum algorithm.

\begin{lstlisting}[language=qpc, float=ht, mathescape=true, escapechar=\%,
label=lst:qpc-example,caption={[Quantum pseudocode for quantum Fourier
transform]Quantum pseudocode for quantum Fourier transform on $d$ qubits.
Quantum circuit for this operation with $d=3$ is presented in
Figure~\ref{fig:qft3}.}]
%{\bf Procedure:} {\sc Fourier}($\underline{a},d$)\\
{\bf Input:} A quantum register $\underline{a}$ with $d$ qubits. Qubits are
numbered from 0 to $d-1$.\\
{\bf Output:} The amplitudes of $\underline{a}$ are Fourier transformed over
$\ \Z_{2^d}$.\\
%
C: assign value to classical variable
$\omega \leftarrow e^{i2\pi/2^d}$
C: perform sequence of gates
for  $i = d-1$ to $i=0$
  for $j = d-1$ to $j=i+1$
  %\underline{\bf if}% $\underline{a_j}$ %\underline{\bf
then}% $\mathcal{R}_{\omega^{2^{d-i-1+j}}}(\underline{a_i})$
   C: number of loops executing phase 
   C: depends on the required accuracy 
   C: of the procedure
  $\mathcal{H}(\underline{a_i})$

C: change the order of qubits
for $j = 0$ to $j=\frac{d}{2}-1$
  $\mathcal{SWAP}(\underline{a_j},\underline{a_{d-a-j}})$
%
%
\end{lstlisting}

Operation $\mathcal{H}(\underline{a_i})$ executes a quantum Hadamard gate on a
quantum register $\underline{a_i}$ and $\mathcal{SWAP}(\underline{a_i},
\underline{a_j})$ performs SWAP gate between $\underline{a_i}$ and
$\underline{a_j}$. Operation $\mathcal{R}_{\phi}(\underline{a_i})$ that executes
a quantum gate $R(\phi)$ is defined as
\begin{equation}
 R(\phi) =     \left(
    \begin{array}{cc}
      1 & 0 \\
      0 & e^{i\phi} \\
    \end{array} 
    \right),
\end{equation}
on the quantum register $\underline{a_i}$. Using conditional contruction
\begin{lstlisting}[language=qpc,mathescape=true,escapechar=\%]
 %\underline{\bf if}% $\underline{a_j}$ %\underline{\bf
then}% $\mathcal{R}_{\phi}(\underline{a_i})$
\end{lstlisting}
it is easy to define controlled phase shift gate (see
Definition~\ref{def:m-toffoli}). Similar construction exists in QCL quantum
programming language described in Section \ref{sec:langs}.

The measurement of a quantum register can be indicated using an assignement
\lstinline [language=qpc,mathescape=true]+$a_j \leftarrow \underline{a_j}$+.

\subsubsection{Quantum programming environment}\label{sec:qenv}
Since the main aim of this paper is to present the advantages and limitations
of high-level quantum programming languages, we need to explain how these
languages are related to quantum random access machine. Thus as the summary of
this section we present the overview of an architecture for quantum programming,
which is based on the QRAM model.

The architecture proposed in \cite{svore-toward,svore06layered} is designed for
transforming a~high-level quantum programming language into the
technology-specific implementation set of operations. This architecture is
composed of four layers:
\begin{itemize}
  \item {\bf High level programming language} providing high-level mechanisms
  for performing useful quantum computation; this language should be independent
  from particular physical implementation of quantum computing.
  
  \item {\bf Compiler of this language} providing architecture independent
  optimisation; also compilation phase can be used to handle quantum error
  correction required to perform useful quantum computation.
  
  \item {\bf Quantum assembly language (QASM)} -- assembly language extended by
  the set of instructions used in the quantum circuit model.
  
  \item {\bf Quantum physical operations language (QCPOL)}, which describes the
  execution of quantum programme in a hardware-dependent way; it includes
  physical operations and it operates on the universal set of gates optimal for
  a given physical implementation.
\end{itemize}

The authors of \cite{svore-toward,svore06layered} do not define a specific
high-level quantum programming language. They point out, however, that existing
languages, mostly based on Dirac notation, do not provide the sufficient level
of abstraction. They also stress, following \cite{bettelliPHD}, that it should
have the basic set of features. We will discuss these basic requirements in
detail in Section \ref{sec:langs}. At the moment quantum assembly language
(QASM) is the most interesting part of this architecture, since it is tightly
connected to the QRAM model.

QASM should be powerful enough for representing high level quantum programming
language and it should allow for describing any quantum circuit. At the same
time it must be implementation-independent so that it could be used to optimise
the execution of the programme with respect to different architectures. 

QASM uses qubits and cbits (classical bit) as basic units of information.
Quantum operations consist of unitary operations and measurement. Moreover, each
unitary operator is expressed in terms of single qubit gates and CNOT gates.

In the architecture proposed in \cite{svore06layered} each single-qubit
operation is stored as the triple of rationals. Each rational multiplied by
$\pi$ represents one of three Euler-angles, which are sufficient to specify
one-qubit operation.

\section{Quantum programming languages}\label{sec:langs}

Quantum algorithms~\cite{shor94algorithms,grover97haystack,grover98any,moscaPHD}
and communication
protocols~\cite{bb84,bennet92communication,brassard05telepathy} are described
using a~language of quantum circuits~\cite{nielsenchuang}. While this method is
convenient in the case of simple algorithms, it is very hard to operate on
compound or abstract data types like arrays or integers using this
notation~\cite{shor04progress,bacon10recent}.

This lack of data types and control structures motivated the development of
quantum pseudocode~\cite{knill96conventions,knill02encyclopedia} and various
quantum programming
languages~\cite{bettelli01toward,mauererMSC,oemerPHD,svore06layered,mlarik07operational,mlarik08semantics}.

Several languages and formal models were proposed for the description of quantum
computation process. The most popular of them is quantum circuit model
\cite{deutsch89networks}, which is tightly connected to the physical operations
implemented in the laboratory. On the other hand the model of quantum Turing
machine is used for analysing the complexity of quantum
algorithms~\cite{bernstein97complexity}.

Another model used to describe quantum computers is Quantum Random Access
Machine (QRAM). In this model we have strictly distinguished the quantum part
performing computation and the classical part, which is used to control
computation. This model is used as a basis for most quantum programming
languages \cite{gay05quantum,unruh06quantum,rudiger07overview}. Among
high-level programming languages designed for quantum computers we can
distinguish imperative and functional languages.

At the moment of writing this paper the most advanced imperative quantum
programming language is Quantum Computation Language (QCL) designed and
implemented by \"Omer \cite{oemerMSC1,oemerMSC2,oemerPHD}. QCL is based on the
syntax of C programming language and provides many elements known from classical
programming languages. The interpreter is implemented using simulation library
for executing quantum programmes on classical computer, but it can be in
principle used as a code generator for classical machine controlling a quantum
circuit.

Along with QCL several other imperative quantum programming languages were
proposed. Notably Q Language developed by
Betteli~\cite{bettelliPHD,bettelli01toward} and libquantum~\cite{libquantum}
have the ability to simulate noisy environment. Thus, they can be used to study
decoherence and analyse the impact of imperfections in quantum systems on the
accuracy of quantum algorithms.

Q Language \cite{bettelli03towards} is implemented as a class library for C++
programming language and libquantum is implemented as a C programming language
library. Q Language provides classes for basic quantum operations like
QHadamard, QFourier, QNot, QSwap, which are derived from the base class Qop. New
operators can be defined using C++ class mechanism. Both Q Language and
libquantum share some limitation with QCL, since it is possible to operate on
single qubits or quantum registers (\ie\ arrays of qubits) only. Thus, they are
similar to packages for computer algebra systems used to simulate quantum
computation~\cite{miszczak05numerical,gawron10extending}.

Concerning problems with physical implementations of quantum computers, it
became clear that one needs to take quantum errors into account when modelling
quantum computational process. Also quantum communication has become very
promising application of quantum information theory over the last few years.
Both facts are reflected in the design of new quantum programming languages.

LanQ developed by Mlna\v{r}\'\i{}k was defined in \cite{lanq,mlarik08semantics}.
It provides syntax based on~C programming language. LanQ provides several
mechanisms such as the creation of a new process by forking and interprocess
communication, which support the implementation of multi-party protocols.
Moreover, operational semantics of LanQ has been defined. Thus, it can be used
for the formal reasoning about quantum algorithms.

It is also worth to mention new quantum programming languages based on
functional paradigm. Research in functional quantum programming languages
started by introducing quantum lambda calculus \cite{tonder04lambda}. It was
introduced in a~form of simulation library for Scheme programming language. QPL
\cite{selinger04towards} was the first functional quantum programming language.
This language is statically typed and allows to detect errors at compile-time
rather than run-time.

A more mature version of QPL is cQPL --- communication capable
QPL~\cite{mauererMSC}. cQPL was created to facilitate the development of new
quantum communication protocols. Its interpreter uses QCL as a backend language
so cQPL programmes are translated into C++ code using QCL simulation library.

Table \ref{tab:qpls-compare} contains the comparison of several quantum
programming languages. It includes the most important features of existing
languages. In particular we list the underlying mathematical model (\ie\ pure or
mixed states) and the support for quantum communication.

\begin{table*}[ht]
    \centering
    \begin{tabular}{l|cccccccc}
    & QCL & Q Language &  QPL & cQPL & LanQ \\\hline
    reference & \cite{oemerMSC1} & \cite{bettelli03towards} &
    \cite{selinger04towards} & \cite{mauererMSC} & \cite{lanq} \\
    implemented & \yes & \yes & \yes & \yes & \yes  \\
    formal semantics & \no & \no &  \yes & \yes & \yes \\
    communication & \no & \no &  \no & \yes & \yes \\
    universal & \yes & \yes & \yes & \yes & \yes  \\
    mixed states & \no & \no &  \yes & \yes & \yes  \\
    \end{tabular}
    \caption{The comparison of
    quantum programming languages with information about implementation and
    basic features. Based on information from \protect\cite{mauererMSC} and
    \protect\cite{lanq}.}
    \label{tab:qpls-compare}
\end{table*}

All languages listed in Table~\ref{tab:qpls-compare} are universal and thus they
can be used to compute any function computable on a quantum Turing machine.
Consequently, all these language provide the model of quantum computation which
is equivalent to the model of a quantum Turing machine.

In this section we compare the selected quantum programming languages and
provide some examples of quantum algorithms and protocols implemented in these
languages. We also describe their main advantages and limitations. We introduce
the basic syntax of three of the languages listed in
Table~\ref{tab:qpls-compare} -- QCL, LanQ and cQPL. This is motivated by the
fact that these languages have a working interpreter and can be used to perform
simulations of quantum algorithms. We introduce basic elements of QCL required
to understand basic programmes. We also compare the main features of the
presented languages.

The main problem with current quantum programming languages is that they tend to
operate on very low-level structures only. In QCL quantum memory can be accessed
using only \texttt{qreg} data type, which represents the array of qubits. In the
syntax of cQPL data type \texttt{qint} has been introduced, but it is only
synonymous for the array of 16 qubits. Similar situation exists in
LanQ~\cite{lanq}, where quantum data types are introduced using
\texttt{q}$n$\texttt{it} keyword, where $n$ represents a dimension of elementary
unit (\eg\ for qubits $n=2$, for qutrits $n=3$). However, only unitary evolution
and measurement can be performed on variables defined using one of these types.

\subsection{Requirements for quantum programming language}
Taking into account QRAM model described in Section \ref{sec:qram-model} we can
formulate basic requirements which have to be fulfilled by any quantum
programming language \cite{bettelli01toward}.

\begin{itemize}
  \item{{\bf Completeness:}} Language must allow to express any quantum circuit
  and thus enable the programmer to code every valid quantum programme written
  as a quantum circuit. 
  
  \item{{\bf Extensibility:}} Language must include, as its subset, the language
  implementing some high level classical computing paradigm. This is important
  since some parts of quantum algorithms (for example Shor's algorithm) require
  nontrivial classical computation.
  
  \item{{\bf Separability:}} Quantum and classical parts of the language should
  be separated. This allows to execute any classical computation on purely
  classical machine without using any quantum resources.
  
  \item{{\bf Expressivity:}} Language has to provide high level elements for
  facilitating the quantum algorithms coding.

  \item{{\bf Independence:}} The language must be independent from any
  particular physical implementation of a quantum machine. It should be possible
  to compile a given programme for different architectures without introducing
  any changes in its source code.
\end{itemize}

As we will see, the languages presented in this Section fulfil most of the above
requirements. The main problem is the \emph{expressivity} requirement.

\subsection{Imperative quantum programming}
First we focus on quantum programming languages which are based on the
imperative paradigm. They include quantum pseudocode, discussed in Section
\ref{sec:qram-model}, Quantum Computation Language (QCL) created by \"Omer
\cite{oemerMSC1,oemerMSC2,oemerPHD} and LanQ developed by Mlna\v{r}\'\i{}k
\cite{mlarik07operational,mlarik08semantics,lanq}

Below we provide an introduction to QCL. It is one of the most popular quantum
programming languages. Next, we introduce the basic elements of LanQ. This
language provides the support for quantum protocols. This fact reflects the
recent progress in quantum communication theory.

\subsubsection{Quantum Computation Language}\label{sec:qcl}
\lstset{language=qcl}
QCL (Quantum Computation Language) \cite{oemerMSC1,oemerMSC2,oemerPHD} is the
most advanced implemented quantum programming language. Its syntax resembles the
syntax of C programming language~\cite{ansic} and classical data types are
similar to data types in~C or Pascal.

The basic built-in quantum data type in QCL is \lstinline{qureg} (quantum
register). It can be interpreted as the array of qubits (quantum bits).

\begin{lstlisting}[caption=Basic operations on quantum registers and subregisters in QCL.]
qureg x1[2]; // 2-qubit quantum register x1
qureg x2[2]; // 2-qubit quantum register x2
H(x1);    // Hadamard operation on x1
H(x2[1]); // and on the second qubit of x2
\end{lstlisting}

QCL standard library provides standard quantum operators used in quantum
algorithms, such as:
\begin{itemize}
\item Hadamard \lstinline{H} and \lstinline{Not} operations on many qubits,
\item controlled not (\lstinline{CNot}) with many target qubits and
\lstinline{Swap} gate,
\item rotations: \lstinline{RotX}, \lstinline{RotY} and \lstinline{RotZ},
\item phase (\lstinline{Phase}) and controlled phase (\lstinline{CPhase}).
\end{itemize}
Most of them are described in Table \ref{tab:gates} in Section \ref{sec:qram-model}.

Since QCL interpreter uses \texttt{qlib} simulation library, it is possible to
observe the internal state of the quantum machine during the execution of 
quantum programmes. The following sequence of commands defines two-qubit
registers \lstinline{a} and \lstinline{b} and executes \lstinline{H} and
\lstinline{CNot} gates on these registers.
\begin{lstlisting}
qcl> qureg a[2];
qcl> qureg b[2];
qcl> H(a);
[4/32] 0.5 |0,0> + 0.5 |1,0> +
 0.5 |2,0> + 0.5 |3,0>
qcl> dump
: STATE: 4 / 32 qubits allocated, 
28 / 32 qubits free
0.5 |0> + 0.5 |1> + 0.5 |2> + 0.5 |3>
qcl> CNot(a[1],b)
[4/32] 0.5 |0,0> + 0.5 |1,0> + 0.5 |2,0> 
+ 0.5 |3,0>
qcl> dump
: STATE: 4 / 32 qubits allocated, 
28 / 32 qubits free
0.5 |0> + 0.5 |1> + 0.5 |2> + 0.5 |3>
\end{lstlisting}
Using \lstinline{dump} command it is possible to inspect the internal state of
a quantum computer. This can be helpful for checking if our algorithm changes
the state of quantum computer in the requested way.

One should note that \lstinline{dump} operation is different from measurement,
since it does not influence the state of quantum machine. This operation can
be realised using simulator only.

\paragraph{Quantum memory management}
Quantum memory can be controlled using quantum types \lstinline{qureg},
\lstinline{quconst}, \lstinline{quvoid} and \lstinline{quscratch}.
Type \lstinline{qureg} is used as a base type for general quantum registers.
Other types allow for the optimisation of generated quantum circuit. The summary
of types defined in QCL is presented in Table \ref{tab:qcl-types}.

\begin{table*}[htp!]
\begin{center}
{  \renewcommand{\arraystretch}{1.2}
\begin{tabular}{|p{2cm}|p{9cm}|p{3.5cm}|}\hline
{\bf Type} & {\bf Description} & {\bf Usage} \\\hline
 \textbf{qureg} & general quantum register & basic type \\
 \textbf{quvoid} & register which has to be empty when operator is called &
target register \\
\textbf{quconst} & must be invariant for all operators used
in quantum conditions & quantum conditions \\
 \textbf{quscratch} & register which has to be empty before and
after the operator is called & temporary registers\\
\hline
\end{tabular}
}
\end{center}
\caption{Types of quantum registers
used for memory management in QCL.}
\label{tab:qcl-types}
\end{table*}

\paragraph{Classical and quantum procedures and functions}
QCL supports user-defined operators and functions known from languages like C or
Pascal. Classical subroutines are defined using \lstinline{procedure} keyword.
Also standard elements, known from C programming language, like looping (\eg\
\lstinline+for i=1 to n { ... }+) and conditional structures (\eg\
\lstinline+if x==0 { ... }+), can be used to control the execution of quantum
and classical elements. In addition to this, it provides two types of quantum
subroutines.

The first type is used for unitary operators. Using it one can define new
operations, which in turn can be used to manipulate quantum data. For example
operator \lstinline{diffuse} defined in Listing~\ref{lst:qcl-diffuse} defines
\emph{inverse about the mean} operator used in Grover's algorithm
\cite{grover97haystack}. This allows to define algorithms on the higher level of
abstraction and extend the library of functions available for a programmer. 

\begin{lstlisting}[language=qcl, float=hb, caption={[Inverse about the mean
operation in QCL]The implementation of the inverse about the mean operation in
QCL \cite{oemerPHD}. Constant \lstinline{pi} represents number $\pi$.
Exclamation mark \lstinline{!} is used to indicate that the interpreter should
use the inverse of a given operator. Operation \lstinline{diffuse} is used in the
quantum search algorithm \cite{grover97haystack}.},label=lst:qcl-diffuse]
operator diffuse(qureg q) {
  H(q);             // Hadamard Transform
  Not(q);           // Invert q
  CPhase(pi,q);     // Rotate if q=1111..
  !Not(q);          // undo inversion
  !H(q);            // undo Hadamard Transform
}
\end{lstlisting}

\begin{figure}[hb!]
  \centering{
  \includegraphics[scale=1.2]{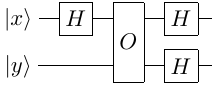}}\\[10pt]

\begin{lstlisting}
operator U(qureg x,qureg y) {
  H(x);
  Oracle(x,y);
  H(x & y);
}

// classical control structure
procedure deutsch() {
  // allocate 2 qubits          
  qureg x[1];                  
  qureg y[1];
  int m;
  // evaluation loop
  {                            
    // initialise machine state
    reset;
    // do unitary computation                     
    U(x,y);
    // measure 2nd register                    
    measure y,m;               
    // value in 1st register valid?  
  } until m==1;

  // measure 1st register which              
  measure x,m;                 
  // contains g(0) xor g(1)
  print "g(0) xor g(1) =",m;
  // clean up   
  reset;                       
}

\end{lstlisting}
\caption{Quantum circuit for Deutsch's algorithm and QCL implementation of this
algorithm (see \protect\cite{oemerPHD} for more examples). Evaluation loop is
composed of preparation (performed by \lstinline{reset} instruction), unitary
evolution (\lstinline{U(x,y)} operator) and measurement. Subroutine
\lstinline{Oracle()} implements function used in Deutsch's algorithm
\protect\cite{deutsch85principle,deutsch92rapid}.}
\label{fig:deutsch-qcl}
\end{figure}

Using subroutines it is easy to describe quantum algorithms. Figure
\ref{fig:deutsch-qcl} presents QCL implementation of Deutsch's algorithm, along
with the quantum circuit for this algorithm. This simple algorithm uses all main
elements of QCL. It also illustrates all main ingredients of existing quantum
algorithms.

The second type of quantum subroutine is called a \emph{quantum
function}. Quantum functions are also called \emph{pseudo-classic
operators}. It can be defined using \lstinline{qufunct} keyword. The subroutine
of type \lstinline{qufunct} is used for all transformations of the form
\begin{equation}
 \ket{n} = \ket{f(n)},
\end{equation}
where $\ket{n}$ is a base state and $f$ is a one-to-one Boolean function.
The example of quantum function is presented in
Listing~\ref{lst:qcl-inc}.

\paragraph{Quantum conditions}
QCL introduces \emph{quantum conditional statements}, \ie\ conditional
constructions where quantum state can be used as a condition.

QCL, as well as many classical programming languages, provides the conditional
construction of the form
\begin{lstlisting}[language=qpc,mathescape=true]
if $be$ then
  $block$
\end{lstlisting}
where $be$ is a Boolean expression and $block$ is a sequence of statements.

QCL provides the means for using quantum variables as conditions. Instead of a
classical Boolean variable, the variable used in condition can be a quantum
register. 
\begin{lstlisting}[mathescape=true,caption=Example of a quantum conditional statement in QCL]
qureg a[2];
qureg b[2];
// the sequence of statements
// ...
// perform CNot if a=$\ket{1\ldots1}$
if a {
  CNot(b[0], b[1]);
}
\end{lstlisting}

In this situation QCL interpreter builds and executes the sequence of $CNOT$
gates equivalent to the above condition. Here register \lstinline{a} is called
\emph{enable register}.

In addition, quantum conditional structures can be used in quantum subroutines.
Quantum operators and functions can be declared as conditional using
\lstinline{cond} keyword. For example
\begin{lstlisting}
// conditional phase gate
extern cond operator Phase(real phi);
// conditional not gate
extern cond qufunct Not(qureg q);
\end{lstlisting}
declares a conditional Phase gate and a controlled $NOT$ gate. Keyword
\lstinline{extern} indicates that the definition of a subroutine is specified in
an external file. The enable register (\ie\ quantum condition) is passed as an
implicit parameter if the operator is used within the body of a quantum
if-statement.

\begin{lstlisting}[float=ht, caption={[Operator for incrementing quantum state
in QCL]Operator for incrementing quantum state in QCL defined as a conditional
quantum function. Subroutine \lstinline{inc} is defined using \lstinline{cond}
keyword and does not require the second argument of type
\lstinline{quconst}. Subroutine \lstinline{cinc} provides equivalent
implementation with explicit-declared enable register.}, label={lst:qcl-inc}]
// increment register
cond qufunct inc(qureg x) { 
  int i;
  for i = #x-1 to 0 step -1 {
  // apply controlled-not from MSB to LSB
    CNot(x[i],x[0::i]);     
  }                         
}

// equivalent implementation 
// with constant enable register
// conditional increment as selection operator
qufunct cinc(qureg x,quconst e) { 
  int i;                          
  for i = #x-1 to 0 step -1 {     
    CNot(x[i],x[0::i] & e);
  }
}
\end{lstlisting}

In the case of \lstinline{inc} procedure, presented in
Listing~\ref{lst:qcl-inc}, the enable register is passed as an implicit
argument. This argument is set by a quantum if-statement and transparently
passed on to all suboperators. As a result, all suboperators have to be
conditional. This is illustrated by the following example \cite{oemerPHD}
\begin{lstlisting}
// counting and control registers
qcl> qureg q[4];qureg e[1];
// prepare test state   
qcl> H(q[3] & e);             
[5/32] 0.5 |0,0> + 0.5 |8,0> + 0.5 |0,1> 
+ 0.5 |8,1>
// conditional increment
qcl> cinc(q,e);               
[5/32] 0.5 |0,0> + 0.5 |8,0> + 0.5 |1,1> 
+ 0.5 |9,1>
// equivalent to cinc(q,e)
qcl> if e { inc(q); }         
[5/32] 0.5 |0,0> + 0.5 |8,0> + 0.5 |2,1> 
+ 0.5 |10,1>
// conditional decrement
qcl> !cinc(q,e);              
[5/32] 0.5 |0,0> + 0.5 |8,0> + 0.5 |1,1> 
+ 0.5 |9,1>
// equivalent to !cinc(q,e);
qcl> if e { !inc(q); }        
[5/32] 0.5 |0,0> + 0.5 |8,0> + 0.5 |0,1> 
+ 0.5 |8,1>
\end{lstlisting}

Finally we should note that a conditional subroutine can be called outside a 
quantum if-statement. In such situation enable register is empty and, as
such, ignored. Subroutine call is in this case unconditional.

\subsubsection{LanQ}
\lstset{language=lanq}
Imperative language LanQ is the first quantum programming language with full
operation semantics specified \cite{lanq}.

Its main feature is the support for creating multipartite quantum protocols.
LanQ, as well as cQPL presented in the next section, are built with quantum
communication in mind. Thus, in contrast to QCL, they provide the features for
facilitating simulation of quantum communication.

Syntax of the LanQ programming language is very similar to the syntax of C
programming language. In particular it supports:
\begin{itemize}
 \item Classical data types: \lstinline{int} and \lstinline{void}.
 \item Conditional statements of the form
\begin{lstlisting}[mathescape=true]
if ( cond ) {
  $\ldots$
} else {
  $\ldots$
}
\end{lstlisting}
\item Looping with \lstinline{while} keyword
\begin{lstlisting}[mathescape=true]
while ( cond ) {
  $\ldots$
}
\end{lstlisting}
\item User-defined functions, for example
\begin{lstlisting}[mathescape=true]
int fun( int i) {
  int res;
  $\ldots$
  return res;
}
\end{lstlisting}
\end{itemize}

\paragraph{Process creation}
LanQ is built around the concepts of process and interprocess communication,
known for example from UNIX operating system. It provides the support for
controlling quantum communication between many parties. The implementation of
teleportation protocol presented in Listing~\ref{lst:lanq-teleport} provides an
example of LanQ features, which can be used to describe quantum communication.

Function \lstinline{main()} in Listing~\ref{lst:lanq-teleport} is responssible
for controlling quantum computation. The execution of protocol is divided into
the following steps:
\begin{enumerate}
    \item Creation of the classical channel for communicating the results of
    measurement: \\ \lstinline{channel[int] c withends [c0,c1];}.
    \item Creation of Bell state used as a quantum
    channel for teleporting a quantum state \\
    (\lstinline+psiEPR aliasfor [psi1, psi2]+); this is accomplished by calling
    external function \lstinline{createEPR()} creating an entangled state.
    \item Instruction \lstinline{fork} executes \lstinline{alice()} function,
    which is used to implement sender; original process continues to run.
    \item In the last step function \lstinline{bob()} implementing a receiver is
    called.
\end{enumerate}

\begin{lstlisting}[language=lanq,  float=ht!, label=lst:lanq-teleport,
caption={[Implementation of teleportation protocol in LanQ]Teleportation
protocol implemented in LanQ \cite{lanq}. Functions
\lstinline{Sigma\_x()}, \lstinline{Sigma\_y()} and \lstinline{Sigma\_z()} are
responsible for implementing Pauli matrices. Function \lstinline{createEPR()}
(not defined in the listing) creates maximally entangled state between parties
--- Alice and Bob. Quantum communication is possible by using the state, which is
stored in a global variable \lstinline{psiEPR}. Function
\lstinline{computeSomething()} (not defined in the listing) is responsible for
preparing a state to be teleported by Alice.}]
void alice(channelEnd[int] c0, 
           qbit auxTeleportState) {
  int i;
  qbit phi;
  // prepare state to be teleported
  phi = computeSomething();
  // Bell measurement
  i = measure (BellBasis, phi, 
                auxTeleportState);
  send (c0, i);
}

void bob(channelEnd[int] c1, 
         qbit stateToTeleportOn) {
  int i;
  i = recv(c1);
  // execute one of the Pauli gates
  // according to the protocol
  if (i == 1) {
    Sigma_z(stateToTeleportOn);
  } else if (i == 2) {
    Sigma_x(stateToTeleportOn);
  } else if (i == 3) {
    Sigma_x(stateToTeleportOn);
    Sigma_z(stateToTeleportOn);
  }
  dump_q(stateToTeleportOn);
}

void main() {
  channel[int] c withends [c0,c1];
  qbit psi1, psi2;
  psiEPR aliasfor [psi1, psi2];

  psiEPR = createEPR();

  c = new channel[int]();
  fork alice(c0, psi1);
  bob(c1, psi2);
}
\end{lstlisting}

\paragraph{Communication}
Communication between parties is supported by providing \lstinline{send}
and \lstinline{recv} keywords. Communication is synchronous, \ie\
\lstinline{recv} delays programme execution until there is a value received from
the channel and \lstinline{send} delays a programme run until the sent value is
received.

Processes can allocate \emph{channels}. It should be stressed that the notion of
channels used in quantum programming is different from the one used in quantum
mechanics. In quantum programming a channel refers to a variable shared between
processes. In quantum mechanics a channel refers to \emph{any quantum
operation}.

Another feature used in quantum communication is variable aliasing. In the
teleportation protocol presented in Listing~\ref{lst:lanq-teleport} the syntax
for variable aliasing
\begin{lstlisting}
  qbit psi1, psi2;
  psiEPR aliasfor [psi1, psi2];
\end{lstlisting}
is used to create quantum state shared among two parties.

\paragraph{Types}
Types in LanQ are used to control the separation between classical and quantum
computation. In particular they are used to prohibit copying of quantum
registers. The language distinguishes two groups of variables \cite[Chapter 5]{lanq}:
\begin{itemize}
    \item Duplicable or non-linear types for representing classical values, \eg\
    \lstinline{bit}, \lstinline{int}, \lstinline{boolean}. The value of a
    duplicable type can be exactly copied.
    \item Non-duplicable or linear types for controlling quantum memory and
    quantum resources, \eg\ \lstinline{qbit}, \lstinline{qtrit} channels and
    channel ends (see example in Listing~\ref{lst:lanq-teleport}). Types from
    this group do not allow for cloning \cite{wootters82single}.
\end{itemize}

One should note that quantum types defined in LanQ are mainly used to check
validity of the program before its run. However, such types do not help to
define abstract operations. As a result, even simple arithmetic operations have
to be implemented using elementary quantum gates, \eg\ using quantum circuits
introduced in~\cite{vedral96arithmetic}.

\subsection{Functional quantum programming -- QPL and cQPL}
During the last few years few quantum programming languages based on functional
programming paradigm have been proposed \cite{selinger04brief}. As we have
already point out, the lack of progress in creating new quantum algorithms is
caused by the problems with operating on complex quantum states. Classical
functional programming languages have many features which allow to clearly
express algorithms \cite{mitchell03concepts}. In particular they allow for
writing better modularised programmes than in the case of imperative programming
languages \cite{hughes89why}. This is important since this allows to debug
programmes more easily and reuse software components, especially in large and
complex software projects.

Quantum functional programming attempts to merge the concepts known from
classical function programming with quantum mechanics. The program in functional
programming language is written as a function, which is defined in terms of
other functions. Classical functional programming languages contain no
assignment statements, and this allows to eliminate side-effects.\footnote{This
is true in \emph{pure functional} programming languages like Haskell.} It means
that function call can have no effect other than to compute its result
\cite{hughes89why}. In particular it cannot change the value of a global
variable.

The first attempts to define a functional quantum programming language were made
by using quantum lambda calculus \cite{tonder04lambda}, which was based on
lambda calculus. For the sake of completeness we can also point out some
research on modelling quantum computation using Haskell programming language
\cite{sabry03modeling,kaczmarczuk03structure}. However, here we focus on
high-level quantum programming languages. Below we present recently proposed
languages QPL and cQPL, which are based on functional paradigm. They aim to
provide mechanisms known from programming languages like Haskell
\cite{hutton07programming} to facilitate the modelling of quantum computation
and quantum communication.

\lstset{language=cqpl}
In \cite{selinger04towards} Quantum Programming Language (QPL) was described and
in \cite{mauererMSC} its extension useful for modelling of quantum communication
was proposed. This extended language was name cQPL -- communication capable QPL.
Since cQPL compiler is also QPL compiler, we will describe cQPL only.

The compiler for cQPL language described in \cite{mauererMSC} is built on the
top of \texttt{libqc} simulation library used in QCL interpreter. As a result,
cQPL provides some features known from QCL.

Classical elements of cQPL are very similar to classical elements of QCL and
LanQ. In particular cQPL provides conditional structures and loops introduced
with \lstinline{while} keyword.
\begin{lstlisting}[escapechar=\%, caption={[Classical elements of cQPL]Classical
control structures in cQPL.}]
new int loop := 10;
while (loop > 5) do {
    print loop;
    loop := loop - 1;
};
if (loop = 3) then {
    print "loop is equal 3";
} else {
    print "loop is not equal 3";
};
\end{lstlisting}

\subsubsection*{Procedures}
Procedures can be defined to improve modularity of programmes.
\begin{lstlisting}[mathescape=true]
proc test: a:int, q:qbit {
  $\ldots$
}
\end{lstlisting}
Procedure call has to know the number of parameters returned by the procedure.
If, for example, procedure \lstinline{test} is defined as above, it is
possible to gather the calculated results
\begin{lstlisting}
new int a1 = 0;
new int cv = 0;
new int qv = 0;
(a1) := call test(cv, qv);
\end{lstlisting}
or ignore them
\begin{lstlisting}
call test(cv, qv);
\end{lstlisting}
In the first case the procedure returns the values of input variables calculated
at the end of its execution.

Classical variables are passed by value \ie\ their value is copied. This is
impossible for quantum variable, since a quantum state cannot be cloned
\cite{wootters82single}. Thus, it is also impossible to assign the value of
quantum variable calculated by procedure.

Note that no cloning theorem requires quantum variables to be \emph{global}.
This shows that in quantum case it is impossible to avoid some effects known
from imperative programming and typically not present in functional programming
languages.

Global quantum variables are used in Listing \ref{lst:cqpl-teleport} to create a
maximally entangled state in a teleportation protocol. Procedure
\lstinline{createEPR(epr1, epr2)} operates on two quantum variables (subsystems)
and produces a Bell state.

\subsubsection*{Quantum elements}
Quantum memory can be accessed in cQPL using variables of type \lstinline{qbit}
or \lstinline{qint}. Basic operations on quantum registers are presented in
Listing \ref{lst:cqpl-q-basic}. In particular, the execution of quantum gates is
performed by using \lstinline{*=} operator.

\begin{lstlisting}[language=cqpl, label=lst:cqpl-basic,
caption={[Basic operations in cQPL]State initialisation and basic gates in
cQPL. Data type \texttt{qbit} represents a single qubit.},
label=lst:cqpl-q-basic]
new qbit q1 := 0;
new qbit q2 := 1;
// execute CNOT gate on both qubits
q1, q2 *= CNot;
// execute phase gate on the first qubit
q1 *= Phase 0.5;
\end{lstlisting}

It should be pointed out that \lstinline{qint} data type provides only a
shortcut for accessing the table of qubits.

Only a few elementary quantum gates are built into the language:
\begin{itemize}
 \item Single qubit gates \texttt{H}, \texttt{Phase} and \texttt{NOT}
implementing basic gates listed in Table~\ref{tab:gates} in
Section~\ref{sec:qram-model}.
 \item \texttt{CNOT} operator implementing controlled negation and
\texttt{FT(n)} operator for $n$-qubit quantum Fourier transform.
\end{itemize}
This allows to simulate an arbitrary quantum computation. Besides, it is
possible to define gates by specifying their matrix elements.

Measurement is performed using \lstinline{measure/then} keywords and
\lstinline{print} command allows to display the value of a variable.

\begin{lstlisting}
measure a then {
  print "a is |0>";
} else {
  print "a is |1>";
};
\end{lstlisting}

In similar manner like in QCL, it is also possible to inspect the value of
a state vector using \lstinline{dump} command.

\subsubsection*{Quantum communication}
The main feature of cQPL is its ability to build and test quantum communication
protocols easily. Communicating parties are described using \emph{modules}. In
analogy to LanQ, cQPL introduces channels, which can be used to send quantum
data. Once again we stress that notion of channels used in cQPL and LanQ is
different from that used in quantum theory. Quantum mechanics introduces
channels to describe allowed physical transformations, while in quantum
programming they are used to describe communication links.

Communicating parties are described by modules, introduced using
\lstinline+module+ keyword. Modules can exchange quantum data (states). This
process is accomplished using \lstinline+send+ and \lstinline+receive+ keywords.

To compare cQPL and LanQ one can use the implementation of the teleportation
protocol. The implementation of teleportation protocol in cQPL is presented in
Listing~\ref{lst:cqpl-teleport}, while the implementation in LanQ is provided in
Listing~\ref{lst:lanq-teleport}.

\begin{lstlisting}[language=cqpl,
label=lst:cqpl-teleport,escapechar=\%,
caption={[Teleportation protocol implemented in cQPL]Teleportation protocol
implemented in cQPL (from \cite{mauererMSC}). Two parties -- Alice and Bob --
are described by modules. Modules in cQPL are introduced using
\lstinline{module} keyword.}]
module Alice {
    proc createEPR: a:qbit, b:qbit {
       a *= H;
       b,a *= CNot; 
       /* b: Control, a: Target */
    } in {
      new qbit teleport := 0;
      new qbit epr1 := 0;
      new qbit epr2 := 0;

      call createEPR(epr1, epr2);
      send epr2 to Bob;

      /* teleport: Control, epr1: Target  
         %\emph{(see: Figure~\ref{fig:teleport})}%  */
      teleport, epr1 *= CNot;

      new bit m1 := 0;
      new bit m2 := 0;
      m1 := measure teleport;
      m2 := measure epr1;

      /* Transmit the classical 
         measurement results to Bob */
      send m1, m2 to Bob;
};

module Bob {
   receive q:qbit from Alice;
   receive m1:bit, m2:bit from Bob;

   if (m1 = 1) then { q *= [[ 0,1,1,0 ]];  /* Apply sigma_x */ };

   if (m2 = 1) then { q *= [[ 1,0,0,-1 ]];  /* Apply sigma_z */};

   /* The state is now teleported */
   dump q;
};
\end{lstlisting}

\section{Summary}
The main goal of this paper is to acquaint the reader with quantum programming
languages and computational models used in quantum information theory. We have
described a quantum Turing machine, quantum circuits and QRAM models of quantum
computation. We have also presented three quantum programming languages --
namely QCL, LanQ and cQPL.

First we should note that the languages presented in this paper provide very
similar set of basic quantum gates and allow to operate only on the arrays of
qubits. Most of the gates provided by these languages correspond to the basic
quantum gates presented in Section \ref{sec:qram-model}. Thus, one can conclude
that the presented languages have the ability to express quantum algorithms
similar to the abilities of a quantum circuit model. 

The biggest advantage of quantum programming languages is their ability to use
classical control structures for controlling the execution of quantum operators.
This is hard to achieve in quantum circuits model and it requires the
introduction of non-unitary operations to this model. In addition, LanQ and cQPL
provide the syntax for clear description of communication protocols.

The syntax of presented languages resembles the syntax of popular classical
programming languages from the C programming language family \cite{ansic}. As
such, it can be easily mastered by programmers familiar with classical
languages. Moreover, the description of quantum algorithms in quantum
programming languages is better suited for people unfamiliar with the notion
used in quantum mechanics.

The main disadvantage of described languages is the lack of quantum data types.
The types defined in described languages are used mainly for two purposes:
\begin{itemize}
    \item To avoid compile-time errors caused by copying of quantum
    registers (cQPL and LanQ).
    \item Optimisation of memory management (QCL).
\end{itemize}
Both reasons are important from the simulations point of view, since they
facilitate writing of correct and optimised quantum programmes. However, these
features do not provide a mechanism for developing new quantum algorithms or
protocols.

\begin{acknowledgments}
The author would like to thank W. Mauerer for providing preliminary version of
his cQPL compiler and acknowledge interesting discussions with P.~Gawron,
B.~\"Omer, I.~Glendinning and H.~Mlna\v{r}\'\i{}k. The author is also grateful
to the anonymous referee for well formulated and interested remarks concerning
the functional quantum programming languages.

This work was supported by the Polish National Science Centre under the grants
number N N516 475440, N N516 481840 and by the Polish Ministry of Science and
Higher Education under the grant number N N519 442339.
\end{acknowledgments}



\end{document}